\documentclass[12pt,draftclsnofoot,onecolumn]{IEEEtran}

\usepackage{cite}
\usepackage{textcomp}
\usepackage{amsfonts}
\usepackage{CJK}
\usepackage{graphicx}
\usepackage{caption}
\usepackage{epstopdf}
\usepackage{subfigure}
\usepackage[cmex10]{amsmath}
\usepackage[amsmath,thmmarks]{ntheorem}
\usepackage{amssymb}
\usepackage{ntheorem}
\usepackage{mathabx}
\usepackage{algorithm,float}
\usepackage{algorithmic}
\usepackage{lipsum}
\usepackage{booktabs}
\usepackage{threeparttable}
\usepackage{multicol}
\usepackage{multirow}
\usepackage{xcolor}
\usepackage{bm}
\usepackage{mathabx}
\usepackage{stmaryrd}
\usepackage{upgreek}
\usepackage{url}
\usepackage{wasysym}
\usepackage{tabularx}
\usepackage[labelsep=none]{caption}

\begin{document}

\title{Intelligent Reflecting Surface Assisted Secure Wireless Communications with Multiple-Transmit and Multiple-Receive Antennas}

\author{Weiheng~Jiang,~\IEEEmembership{Member, IEEE},
        Yu~Zhang,
        Jinsong~Wu,~\IEEEmembership{Senior Member, IEEE},
        Wenjiang~Feng,
        Yi~Jin
\thanks{W. Jiang, Y. Zhang and W. Feng are with the School of Microelectronic and Communication Engineering, Chongqing University, Chongqing, 400044 China e-mail: \{whjiang, zhangyu2014, fengwj\}@cqu.edu.cn.}% <-this % stops a space
\thanks{J. Wu is with the Department of Electrical Engineering, Universidad de Chile, Santiago, 8370451 Chile e-mail: wujs@ieee.org.}% <-this % stops a space
\thanks{Y. Jin is with the Xi'an Branch of China Academy of Space Technology, Xi'an, 710100 China email: john.0216@163.com}
        }

\markboth{Journal of \LaTeX\ Class Files,~Vol.~14, No.~8, August~2015}%
{Shell \MakeLowercase{\textit{et al.}}: Bare Demo of IEEEtran.cls for IEEE Journals}

\maketitle

\begin{abstract}
In this paper, we propose intelligent reflecting surfaces (IRS) assisted secure wireless communications with multi-input and multi-output antennas (IRS-MIMOME). The considered scenario is an access point (AP) equipped with multiple antennas communicates with a multi-antenna enabled legitimate user in the downlink at the present of an eavesdropper configured with multiple antennas. Particularly, the joint optimization of the transmit covariance matrix at the AP and the reflecting coefficients at the IRS to maximize the secrecy rate for the IRS-MIMOME system is investigated, with two different assumptions on the phase shifting capabilities at the IRS, i.e., the IRS has the continuous reflecting coefficients and the IRS has the discrete reflecting coefficients. For the former case, due to the non-convexity of the formulated problem, an alternating optimization (AO)-based algorithm is proposed, i.e., for given the reflecting coefficients at the IRS, the successive convex approximation (SCA)-based algorithm is used to solve the transmit covariance matrix optimization, while given the transmit covariance matrix at the AP, alternative optimization is used again in individually optimizing of each reflecting coefficient at the IRS with other fixed reflecting coefficients. For the individual reflecting coefficient optimization, the close-form or an interval of the optimal solution is provided. Then, the proposed algorithm is extended to the discrete reflecting coefficient model at the IRS. Finally, some numerical simulations have been done to demonstrate that the proposed algorithm outperforms  other benchmark schemes.
\end{abstract}


\IEEEpeerreviewmaketitle

\section{Introduction}

\IEEEPARstart{D}{ue} to the broadcast nature of the wireless media, wireless communications are vulnerable to eavesdropping. In order to provide the wireless communications with sound and solid security, physical layer security based technologies, such as the artificial noise (AN), cooperative jamming (CJ) and friendly jamming, have been studied for the recent years \cite{DBLP:journals/comsur/ChenNGC17, DBLP:journals/comsur/WangBZH19}. However, these techniques only focus on the signal processing at the transceiver to adaptive the changes of the wireless environments, but cannot eliminate the negative effects caused by the uncontrollable electromagnetic wave propagation environments \cite{DBLP:journals/cm/LiaskosNTPIA18, YangA}. Meanwhile, recently, a new technology following the development of the Micro-Electro-Mechanical Systems (MEMS) named as intelligent reflecting surfaces (IRS) has been proposed, which can reconfigure the wireless propagation environment via software-controlled reflection \cite{DBLP:journals/cm/LiaskosNTPIA18, YangA, DBLP:journals/corr/abs-1903-08925} and shows tremendous potentials in enhancing the wireless communication performance, such as the transmission rate and security, with low cost and significant performance gain, and has received considerable attentions.

One the one hand, for the IRS assisted wireless communications, in \cite{DBLP:conf/globecom/WuZ18}, the problem of jointly optimizing the access point (AP) active beamforming and IRS passive beamforming with AP transmission power constraint to maximize the received signal power for one pair of transceivers was discussed. Based on the semidefinite relaxation and the alternate optimization, both the centralized algorithm and distributed algorithm were proposed therein. The work \cite{DBLP:journals/twc/WuZ19} extended the previous work to the multi-users scenario but with the individual signal-to-noise ratio (SNR) constraints, where the joint optimization of the AP active beamforming and IRS passive beamforming was discussed to minimize the total AP transmission power, and two suboptimal algorithms with different performance-complexity tradeoff were presented. Huang et al. considered the IRS-based multiple-input single-output (MISO) downlink multi-user communications for an outdoor environment, where \cite{DBLP:conf/icassp/HuangZDY18} studied optimizing the base station (BS) transmission power and IRS phase shift with BS transmission power constraint and user signal-to-interference-and-noise-ratio (SINR) constraint to maximize sum system rate. Since the formulated resource allocation problem is non-convex,  Majorization-Minimization (MM) and alternating optimization (AO) was jointly used, and the convergence of this algorithm was analyzed. Different from the continuous phase shift assumption of the IRS reflecting elements in existing studies, \cite{DBLP:conf/icassp/WuZ19} considered that each IRS reflecting element can only achieve discrete phase shift and the joint optimization of the multi-antenna AP beamforming and IRS discrete phase shift was discussed under the same scenario as \cite{DBLP:conf/globecom/WuZ18}. Then the performance loss caused by the IRS discrete phase shift was quantitatively analyzed via comparing with the IRS continuous phase shift. It is surprised that, the results have shown that as the number of IRS reflecting elements approaches infinity, the system can obtain the same square power gain as IRS with continuous phase shift, even based on 1-bit discrete phase shift. Furthermore, \cite{DBLP:journals/corr/abs-1810-06934} and \cite{DBLP:conf/globecom/HuangAZDY18} discussed the joint AP power allocation and IRS phase-shift optimization to maximize system energy and spectrum efficiency, where the user has a minimum transmission rate constraint and the AP has a total transmit power constraint. Due to the presented problem is non-convex, the gradient descent based AP power allocation algorithm and fractional programming (FP) based IRS phase shift algorithm were proposed therein. For the IRS assisted wireless communication system, Han and Tang et al. \cite{DBLP:journals/tvt/HanTJWM19} analyzed and obtained a compact approximation of system ergodic capacity and then, based on statistical channel information and approximate traversal capacity, the optimal IRS phase shift was proved. The authors also derived the required quantized bits of the IRS discrete phase shift system to obtain an acceptable ergodic capacity degradation. In \cite{DBLP:journals/corr/abs-1904-10136}, a new IRS hardware architecture was presented and then, based on compressed sensing and deep learning, two reflection beamforming methods were proposed with different algorithm complexity and channel estimation training overhead. Similar to \cite{DBLP:journals/corr/abs-1904-10136}, Huang and Debbah et al. \cite{DBLP:conf/spawc/HuangAYD19} proposed a deep learning based algorithm to maximize the received signal strength for IRS-assisted indoor wireless communication environment. Some recently studies about the IRS assisted wireless communications could be found in \cite{DBLP:conf/icc/TanSJP16, DBLP:conf/infocom/TanSKJ18, DBLP:journals/corr/abs-1907-03133, DBLP:journals/corr/abs-1907-06002}, and they were focused on the IRS assisted millimeter band or non-orthogonal multiple access (NOMA) based wireless communications.

On the other hand, for the IRS assisted secure wireless communications, in \cite{DBLP:journals/corr/abs-1904-09573}, the authors studied the problem in jointly optimizing the beamforming at the transmitter and the IRS phase shifts to maximize the system secrecy rate, based on the block coordinate descent (BCD) and the MM techniques, two suboptimal algorithms were proposed to solve the resulted non-convex optimization problem for small- and large-scale IRS, respectively. In \cite{DBLP:journals/access/ChenLPG19}, Chen and Liang studied the minimum-secrecy-rate maximizing problem for a downlink MISO broadcast system, based on the AO and the path-following (PF) algorithm, an iterative algorithm was proposed for the joint optimization problem. In addition, the authors also extended the proposed approach to the case with discrete reflecting coefficients at the IRS. To maximize the MISO system secrecy rate subject to the source transmit power constraint and the unit modulus constraints imposed on the phase shifts at the IRS, \cite{DBLP:journals/icl/ShenXGHZ19} proposed an AO algorithm for the scenario that the eavesdropper is configured with single antenna, then the study was extended to the scenario where the eavesdropper is equipped with multiple antennas. \cite{DBLP:journals/corr/abs-1909-00629} investigated the secure transmission framework with an IRS to minimize the system energy consumption in cases of rank-one and full-rank AP-IRS links. In particular, since the beamforming vector and phase shift design are independent in the rank-one channel model, thus a closed-form expression of beamforming vector was derived. However, since beamforming and phase shift depend on each other in the full-rank model, then an eigenvalue-based algorithm for conventional wiretap channel was used to obtain beamforming vector. Different from \cite{DBLP:journals/corr/abs-1904-09573, DBLP:journals/access/ChenLPG19, DBLP:journals/icl/ShenXGHZ19,DBLP:journals/corr/abs-1909-00629}, \cite{DBLP:journals/wcl/CuiZZ19} considered the scenario that the eavesdropping channel is stronger than the legitimate channel and they are also highly correlated in space, then to maximize the secrecy rate of the legitimate communication link, an algorithm based on the AO and semidefinite relaxation was proposed. Moreover, in \cite{DBLP:journals/corr/abs-1907-12839} and \cite{DBLP:journals/corr/abs-1907-03085}, for the IRS assisted MISO secure communications with AN transmission at the transmitter, an alternate optimization algorithm to jointly optimize active beamforming, AN interference vector and reflection beamforming with the goal of maximizing system secrecy rate was presented. The difference between these two papers is that, \cite{DBLP:journals/corr/abs-1907-12839} focused on the scenario with a single legitimate user and multiple eavesdroppers, while \cite{DBLP:journals/corr/abs-1907-03085} considered the scenario with multiple legitimate users but single eavesdropper.

Although lots of research works have been done for the IRS assisted secure communications, they all have assumed that the legitimate receiver is equipped with only one antenna \cite{DBLP:journals/corr/abs-1904-09573, DBLP:journals/access/ChenLPG19, DBLP:journals/icl/ShenXGHZ19, DBLP:journals/corr/abs-1909-00629, DBLP:journals/wcl/CuiZZ19, DBLP:journals/corr/abs-1907-12839, DBLP:journals/corr/abs-1907-03085}. However, in order to further improve the communication performance of the mobile users in the next generation wireless local networks (WLANs) such as the IEEE 802.11ax, or the fifth generation (5G) mobile communication networks, multi-antenna enabled mobile device designs have been widely adopted in the current mobile terminals, such as the Phones, laptops and the tablets. Therefore, it is necessary to study the IRS assisted secure communications with multiple-transmit and multiple-receive antennas enabled networks. In this paper, IRS assisted secure communications with multiple-transmit and multiple-receive antennas are studied, where, an AP equipped with multiple antennas has the secure communications demands with a multiple-antennas enabled legitimate user in the downlink at the present of an eavesdropper configured with multiple antennas, referring to it as the IRS assisted multi-input, multi-output, multi-eavesdropper (IRS-MIMOME) system. Particularly, we discuss the joint optimization of the transmit covariance matrix at the AP and the reflecting coefficients at the IRS to maximize the secrecy rate for the IRS-MIMOME system, with two different assumptions on the phase shifting capabilities at the IRS, i.e., the IRS has the continuous reflecting coefficients and the IRS has the discrete reflecting coefficients. For the former case, due to the non-convexity of the formulated problem, an AO based algorithm is proposed, i.e., for given the reflecting coefficients at the IRS, the successive convex approximation (SCA)-based algorithm is used to solve the transmit covariance matrix optimization, while given the transmit covariance matrix at the AP, alternative optimization is used again in the individually optimizing of each reflecting coefficient at the IRS with fixing the other reflecting coefficients. For the individual reflecting coefficient optimization, the close-form or an interval of the optimal solution is provided. Then, the overall algorithm was extended to the discrete reflecting coefficient model at the IRS. Finally, some numerical simulations have been done to demonstrate the performance of the proposed algorithms.

The rest parts of this paper are organized as follows. In Section \ref{section_2}, the system model and the considered optimization problem are presented. In Section \ref{section_3}, we discuss and solve the formulated optimization problem, and an AO based algorithm is proposed. The simulation results are presented in Section \ref{section_4} and then we conclude this paper.

\emph{Notation:} We use uppercase boldface letters for matrices and lowercase boldface letters for vectors. $(\bullet)^T$, $(\bullet)^*$, and $(\bullet)^H$ denote the transpose, conjugate, and conjugate transpose, respectively. $Tr(\bullet)$ and $\mathbb{E}\{\bullet\}$ stand for the trace of a matrix and the statistical expectation for random variables, respectively. $\mathbf{A} \succcurlyeq 0$ and $\mathbf{A} \succ 0$ indicate that $\mathbf{A}$ are positive semidefinite and positive definite matrix. $\mathbf{I}$ and $(\bullet)^{-1}$ denote the identity matrix with appropriate size and the inverse of a matrix, respectively. $|\bullet|$, $arg(\bullet)$ and $\Re \{ \bullet \}$ stand for the absolute value, the argument and the real part of a complex number, respectively, whereas $\det(\mathbf{A})$ denotes the determinant of $\mathbf{A}$. The notation $diag(\bullet)$ represents a diagonal matrix where the diagonal elements are from a vector, and $[\bullet]^+$ represents $\max( 0, \bullet)$.

\section{System Model and The Problem}\label{section_2}
In this section, firstly, we present the system model of the IRS assisted secure communications with multiple antennas at both the legitimate transceiver and the eavesdropper, referring to it as the IRS-MIMOME system. Then, we illustrate the IRS reflecting model and signal model for our considered system. Finally, we formulate the discussed optimization problem.

\subsection{System Model}
\begin{figure}[tbp]
    \centering
    \includegraphics[scale=0.8]{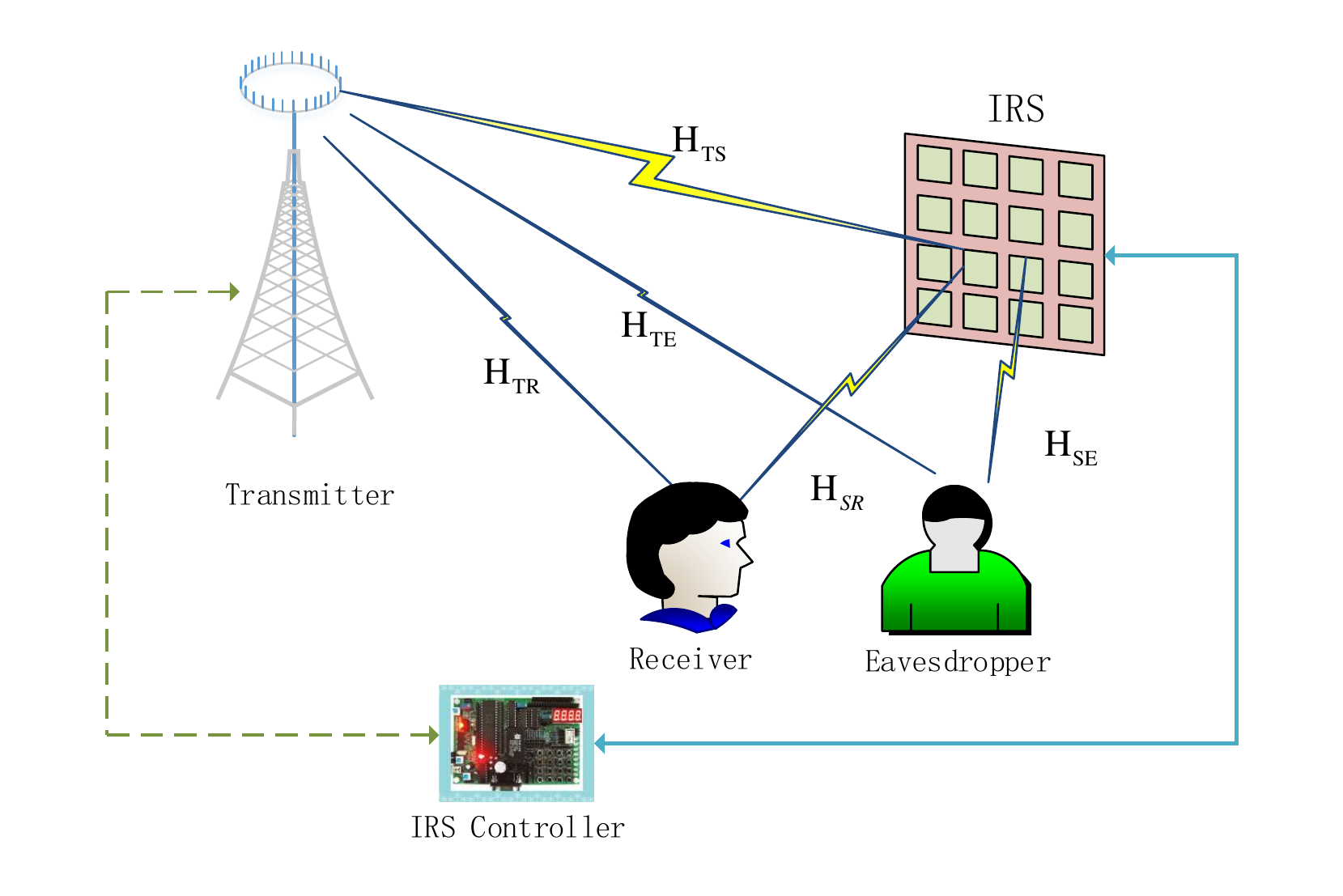}
    \caption{\quad System Model.}
    \label{fig1}
\end{figure}

Consider the IRS assisted MIMOME system, as shown in Fig.~\ref{fig1}, where an AP equipped with $N_T$ antennas serves a legitimate user at the present of an eavesdropper. Both the legitimate user and the eavesdropper are equipped with multi-antenna and the number of the antennas at these two users are $N_R$ and $N_E$, respectively. In addition, an IRS composed of $M$ passive elements is installed on a surrounding wall to assist the secure communications between the AP and the legitimate user. The IRS has a smart controller, who has the capability of dynamically adjusting the phase shift of each reflecting element based on the propagation environment learned through periodic sensing \cite{DBLP:conf/globecom/WuZ18}. In particular, the IRS controller coordinates the switching between two working modes, i.e., receiving mode for environment sensing (e.g.,  channel state information (CSI) estimation) and reflecting mode for scattering the incident signals from the AP \cite{DBLP:journals/iet-com/SubrtP12}.
%²åÈëϵͳģÐÍͼ

\subsection{IRS Reflecting Coefficient Model}
Following \cite{DBLP:journals/corr/abs-1910-07156}, the phase shift matrix of the IRS can be defined as $\mathbf{\mathbf{\Theta}} = diag (\boldsymbol{\theta}) \in {{\mathbb{C}}^{M \times M}}$, where $\boldsymbol{\theta}=[\theta_1,\theta_2,\cdots,\theta_M]^T \in {{\mathbb{C}}^{M \times 1}}$ and $\theta_m \in \boldsymbol{\Phi}$ for $1 \le m \le M$, and $diag(\bullet )$ denotes a diagonal matrix whose diagonal elements are given by the corresponding vector and $\boldsymbol{\Phi}$ denotes the set of reflecting coefficients of the IRS. In this paper, two different sets of reflecting coefficients are considered as below.

\subsubsection{Continuous Reflecting Coefficients} That is, the reflecting coefficient with the constant amplitude and continuous phase shift is characterized as
%ÀëÉ¢ÏàÒÆÄ£ÐÍ
\begin{equation}\label{eq1}
{\boldsymbol{\Phi} _1} = \left\{ {{\theta_n}\left| {{\theta_n} = {e^{j{\varphi _n}}},{\varphi _n} \in \left[ {0,2\pi } \right)} \right.} \right\}.
\end{equation}

\subsubsection{Discrete Reflecting Coefficients} In this model, the reflecting coefficient has constant amplitude but discrete phase shift and is defined as
%Á¬ÐøÏàÒÆÄ£ÐÍ
\begin{equation}\label{eq2}
{\boldsymbol{\Phi} _2} = \left\{ {{\theta _n}\left| {{\theta _n} = {e^{j{\varphi _n}}},{\varphi _n} \in \left\{ {0,\frac{{2\pi }}{Q},...,\frac{{2\pi (Q - 1)}}{Q}} \right\}} \right.} \right\},
\end{equation}
where $Q$ is the number of quantized reflection coefficient values of the element of the IRS.

Note that, due to the limitations of the hardware, the realization of the continuous reflecting model $\boldsymbol{\Phi} _1$ is difficult or even impossible \cite{DBLP:journals/access/ChenLPG19}. Therefore, the discrete model $\boldsymbol{\Phi} _2$ is more practical from the perspective of application. However, the continuous reflecting model is still discussed herein for the obtained performance can be regarded as the upper bound of the system. Furthermore, our algorithm for the discrete model is based on the algorithm of the continuous model.

\subsection{Signal Model}
For our considered system, as \cite{DBLP:conf/globecom/WuZ18}, the signals that are reflected by the IRS multi-times are ignored due to significant path loss. Therefore, combined with IRS reflecting coefficient model, the signals received at the legitimate user and the eavesdropper can be expressed as
%ÊÚȨÓû§ÐźÅÄ£ÐÍ
\begin{equation}\label{eq3}
{\boldsymbol{y}}_R = {\mathbf{H}_{TR}}{\boldsymbol{x}} + {\mathbf{H}_{SR}}{\mathbf{\mathbf{\Theta}}}{\mathbf{H}_{TS}}{\boldsymbol{x}}+{\boldsymbol{n}}_R={\mathbf{G}_{TR}(\mathbf{\mathbf{\Theta}})}{\boldsymbol{x}}+{\boldsymbol{n}}_R,
\end{equation}
%ÇÔÌýÕßÐźÅÄ£ÐÍ
\begin{equation}\label{eq4}
{\boldsymbol{y}}_E = {\mathbf{H}_{TE}}{\boldsymbol{x}} + {\mathbf{H}_{SE}}{\mathbf{\Theta}}{\mathbf{H}_{TS}}{\boldsymbol{x}}+{\boldsymbol{n}}_E={\mathbf{G}_{TE}(\mathbf{\Theta})}{\boldsymbol{x}}+{\boldsymbol{n}}_E,
\end{equation}

%%ÕâÒ»²¿·Ö±íÊö¸ÄÒ»ÏÂ˳Ðò£¬ÏÈдÐŵÀ£¬ÒÔ¼°µÈ¼ÛÐŵÀ£¬
\noindent where $\mathbf{H}_{TR}\in{{\mathbb{C}}^{{N_R} \times {N_T}}}$ and $\mathbf{H}_{TE}\in{{\mathbb{C}}^{{N_E} \times {N_T}}}$ represent the complex baseband channels from AP to the legitimate user and the eavesdropper, respectively, $\mathbf{H}_{SR}\in{{\mathbb{C}}^{{N_R} \times M}}$ and $\mathbf{H}_{SE}\in{{\mathbb{C}}^{{N_E} \times M}}$ denote the complex baseband channels from the IRS to the legitimate user and the eavesdropper, respectively, and $\mathbf{H}_{TS}\in{{\mathbb{C}}^{M \times {N_T}}}$ defines the complex baseband channel from AP to the IRS. $\mathbf{G}_{TR}\left(\mathbf{\Theta}\right)={\mathbf{H}_{TR}}+{\mathbf{H}_{SR}}\mathbf{\Theta}{\mathbf{H}_{TS}}$ and $\mathbf{G}_{TE}\left(\mathbf{\Theta}\right)={\mathbf{H}_{TE}}+{\mathbf{H}_{SE}}\mathbf{\Theta}{\mathbf{H}_{TS}}$ are used to characterize the equivalent channel from AP to the legitimate user and the eavesdropper, respectively. ${{\boldsymbol{n}}_R}\sim {CN\left(0,{\sigma _R^2{\mathbf{I}_{{N_R}}}}\right)}$ and ${{\boldsymbol{n}}_E}\sim {CN\left(0,{\sigma _E^2{\mathbf{I}_{{N_E}}}}\right)}$ denote the independent circularly symmetric complex Gaussian (CSCG) noise vectors at the legitimate user and the eavesdropper, respectively. In which, $\sigma _R^2$ and $\sigma _E^2$ denote the average noise power at the legitimate user and the eavesdropper, respectively. ${\mathbf{I}_{{N_R}}}$ and ${\mathbf{I}_{{N_E}}}$ represent the identity matrix with ${N_R} \times {N_R}$ and ${N_E} \times {N_E}$ dimensions, respectively. As in  \cite{DBLP:journals/corr/abs-1907-05558}, the quasi-static flat-fading channel model is adopted herein and all the CSI are perfectly known at the AP.

For the above established IRS-MIMOME system, following \cite{DBLP:journals/icl/ZhangHLY14}, \cite{DBLP:journals/ett/Telatar99} and \cite{DBLP:journals/icl/Park16}, we know that its achievable secrecy rate is
%°²È«ËÙÂʶ¨Òåʽ
\begin{equation}\label{eq5}
{R_{sec }} = {\left[ {{R_R} - {R_E}} \right]^{\rm{ + }}},
\end{equation}
\noindent where ${\left[ x \right]^ + } = \max \left( {0,x} \right)$. And $R_R$ and $R_E$ represent the achievable transmission rates from AP to the legitimate user and from AP to the eavesdropper, respectively, and they are defined as follows,
%ÊÚȨÓû§ËÙÂÊ
\begin{equation}\label{eq6}
{R_R}={\log_2}\det\left( {\mathbf{I}_{N_R}}+{\frac{1}{{\sigma _R^2}}{\mathbf{G}_{TR}(\mathbf{\Theta})}{\mathbf{Q}_s}\mathbf{G}_{TR}^H(\mathbf{\Theta})}\right),
\end{equation}
%ÇÔÌýÕßËÙÂÊ
\begin{equation}\label{eq7}
{R_E}={\log _2}\det\left( {\mathbf{I}_{N_E}}+{\frac{1}{{\sigma _E^2}}{\mathbf{G}_{TE}(\mathbf{\Theta})}{\mathbf{Q}_s}\mathbf{G}_{TE}^H(\mathbf{\Theta})}\right),
\end{equation}
where ${\mathbf{Q}_s}=\mathbb{E}\left\{{\boldsymbol{x}}{\boldsymbol{x}}^H\right\}\in{{\mathbb{C}}^{{N_T} \times {N_T}}}$ is the transmit signal covariance matrix at the AP and ${\mathbf{Q}_s}\succcurlyeq 0$. Hence, the achievable secrecy rate for the legitimate user is characterized by
%°²È«ËÙÂʱí´ïʽ
\begin{equation}\label{eq8}
\begin{split}
R_{sec}&=\left[{\log_2}\det\left(\mathbf{I}_{N_R} + \frac{1}{\sigma_R^2}\mathbf{G}_{TR}(\mathbf{\Theta})\mathbf{Q}_s\mathbf{G}^{H}_{TR}(\mathbf{\Theta})\right)\right.
    \\&\left.-{\log_2}\det\left({\mathbf{I}_{N_E}} + \frac{1}{{\sigma _E^2}} \mathbf{G}_{TE}(\mathbf{\Theta})\mathbf{Q}_s\mathbf{G}^{H}_{TE}(\mathbf{\Theta})\right)\right]^+.
\end{split}
\end{equation}
Note that dropping the operator $[\bullet]^+$ has no impact on the optimization of the secrecy rate, thus this operator is removed in the sequel analysis.

\subsection{Problem Formulation}
As mentioned earlier, in this paper, we discuss the joint optimization of the transmit covariance matrix at the AP and the reflection coefficients at the IRS to maximize the system secrecy rate subjected to the transmit power constraint at the AP and the reflection coefficient constraint at the IRS. Thus we have the following optimization problem OP1,
%ÓÅ»¯ÎÊÌâÄ£ÐÍ
\begin{equation}\label{eq9}
\begin{split}
\mathop {\max}\limits_{{\mathbf{Q}_s},\mathbf{\Theta} }\;R_{sec}&={\log_2}\det\left(\mathbf{I}_{N_R} + \frac{1}{\sigma_R^2}\mathbf{G}_{TR}(\mathbf{\Theta})\mathbf{Q}_s\mathbf{G}^{H}_{TR}(\mathbf{\Theta})\right)\\
    &-{\log_2}\det\left({\mathbf{I}_{N_E}} + \frac{1}{{\sigma _E^2}} \mathbf{G}_{TE}(\mathbf{\Theta})\mathbf{Q}_s\mathbf{G}^{H}_{TE}(\mathbf{\Theta})\right)\\
s.t.\;&C1:Tr({\mathbf{Q}_s}) \le {P_{\max }}\\
    &C2:{\mathbf{Q}_s} \succcurlyeq 0\\
    &C3:{\theta _m}\in\boldsymbol{\Phi}_i, m = 1,...,M, i =1,2.
\end{split}.
\end{equation}
In which, C1 characterizes the total transmit power constraint at the AP, C2 defines the positive semi-define constraint on transmit covariance matrix, and C3 represents the IRS reflecting coefficient model. It is obvious that OP1 is a non-convex nonlinear programming with non-convex objective function and the uni-modular constraint on each reflection coefficient $\theta_m$, which makes it difficult to be solved. Therefore, in the sequel, we pursue the suboptimal approach to handle OP1.

\section{Alternating Optimization based Joint optimization Algorithm}\label{section_3}
In this section, a suboptimal algorithm is proposed to solve OP1. As aforementioned that, our formulated problem OP1 is a non-convex nonlinear programming. However, our analysis indicates that, given the reflecting coefficients at the IRS and by leveraging the SCA \cite{DBLP:journals/tvt/CumananDSTL14}, a convex approach can be used to solve the transmit covariance matrix optimization at the AP, while for given the transmit covariance matrix at the AP, we can use the alternative optimization to find the suboptimal solution for the reflecting coefficients at the IRS. Based on that, we present an alternative suboptimal algorithm for OP1. In addition, we also discuss the extension of the algorithm to the case with discrete reflecting coefficients at the IRS at the end of this section.

\subsection{Optimization of the transmit covariance matrix}
In this subsection, we discuss the transmit covariance matrix optimization at the AP for given the reflecting coefficients at the IRS. Hence, we have the following problem OP2,
%·¢ÉäЭ·½²îÓÅ»¯×ÓÎÊÌâ
\begin{equation}\label{eq10}
\begin{split}
\mathop {\max }\limits_{\mathbf{Q}_s}{R_{sec}}\left({\mathbf{Q}_s}\right) &={R_R}\left({\mathbf{Q}_s}\right)-{R_E}\left({\mathbf{Q}_s}\right)\\
&={{\log}_2}\det \left( {\mathbf{I}_{N_R}}+\frac{1}{\sigma _R^2}{\mathbf{G}_{TR}}{\mathbf{Q}_s}{\mathbf{G}_{TR}^H} \right)\\
&-{{\log}_2}\det \left( {\mathbf{I}_{N_E}}+\frac{1}{\sigma _E^2}{\mathbf{G}_{TE}}{\mathbf{Q}_s}{\mathbf{G}_{TE}^H} \right),\\
s.t.&C1:Tr({\mathbf{Q}_s}) \le {P_{\max }},\\
    &C2:{\mathbf{Q}_s} \succcurlyeq 0.
\end{split}
\end{equation}
Herein, given the reflection coefficient matrix $\mathbf{\Theta}$ at the IRS, $\mathbf{G}_{TR}(\mathbf{\Theta})$ and $\mathbf{G}_{TE}(\mathbf{\Theta})$ are simplistically denoted as $\mathbf{G}_{TR}$ and $\mathbf{G}_{TE}$, respectively. One may note that, now, the formulated problem OP2 is the secrecy rate maximization problem for the MIMOME system which has been discussed in \cite{DBLP:journals/tvt/CumananDSTL14}, \cite{Steinwandt2014SECRECY} and \cite{DBLP:journals/tsp/FakoorianS13a}, and various algorithms have been proposed therein. In this paper, following \cite{DBLP:journals/tvt/CumananDSTL14}, the SCA-based suboptimal algorithm is used to handle OP2. And the key point is to obtain a tight concave lower bound of ${R_{sec}}\left({\mathbf{Q}_s}\right)$, which can be achieved by retaining the concave part ${R_R}\left({\mathbf{Q}_s}\right)$ in (\ref{eq10}) and linearizing the concave function ${R_E}\left({\mathbf{Q}_s}\right)$ \cite{DBLP:journals/tsp/ScutariFL17, DBLP:journals/tsp/ScutariFLSS17}. That is, at ${\mathbf{\tilde{Q}}}_s$, we have the concave approximation of ${R_{sec}}\left({\mathbf{Q}_s}\right)$ as follows,
%½üËÆ°¼Ï½ç
\begin{equation}\label{eq11}
\begin{split}
{R_{sec }}\left( {\mathbf{Q}_s} \right)&={{\log }_{2}}\det \left( {{\mathbf{I}}_{{{N}_{R}}}}+\frac{1}{\sigma _{R}^{2}}{{\mathbf{G}}_{TR}}{{\mathbf{Q}}_{s}}\mathbf{G}_{TR}^{H} \right)-{{\log }_{2}}\det \left( {{\mathbf{I}}_{{{N}_{E}}}}+\frac{1}{\sigma _{E}^{2}}{{\mathbf{G}}_{TE}}{{\mathbf{Q}}_{s}}\mathbf{G}_{TE}^{H} \right)\\
&\simeq {{\log }_{2}}\det \left( {{\mathbf{I}}_{{{N}_{R}}}}+\frac{1}{\sigma _{R}^{2}}{{\mathbf{G}}_{TR}}{{\mathbf{Q}}_{s}}\mathbf{G}_{TR}^{H} \right)-{{\log }_{2}}\det \left( {{\mathbf{I}}_{{{N}_{E}}}}+\frac{1}{\sigma _{E}^{2}}{{\mathbf{G}}_{TE}}{{{\mathbf{\tilde{Q}}}}_{s}}\mathbf{G}_{TE}^{H} \right)\\
&-Tr\left[ \frac{1}{ln2}\frac{1}{\sigma _{E}^{2}}{{\mathbf{W}^{-1}_{E,{\mathbf{\tilde{Q}}_s}}}}{{\mathbf{G}}_{TE}}{{\mathbf{Q}}_{s}}\mathbf{G}_{TE}^{H} \right]+Tr\left[ \frac{1}{ln2}\frac{1}{\sigma _{E}^{2}}{{\mathbf{W}^{-1}_{E,{\mathbf{\tilde{Q}}_s}}}}{{\mathbf{G}}_{TE}}{{{\mathbf{\tilde{Q}}}}_{s}}\mathbf{G}_{TE}^{H} \right]\\
&\buildrel \Delta \over = {{{\tilde{R}}}_{sec }}\left( {{\mathbf{Q}}_{s}}\left| {{{\mathbf{\tilde{Q}}}}_{s}} \right. \right),\\
\end{split}
\end{equation}
where $\mathbf{W}_{E,{\mathbf{\tilde{Q}}_s}}={\mathbf{I}_{N_E}}+\frac{1}{\sigma _E^2}{\mathbf{G}_{TE}} {{\mathbf{\tilde{Q}}}_s}\mathbf{G}_{TE}^{H}$. Based on the above approximation and given ${\mathbf{\tilde{Q}}}_s$, the problem OP2 can be transformed  into the following formulation,
%ת»¯ºóµÄ×ÓÎÊÌâ
\begin{equation}\label{eq12}
\begin{split}
\mathop {\max }\limits_{\mathbf{Q}_s}\;&{\tilde{R}_{sec }}\left( {\mathbf{Q}_s}\left| {\mathbf{\tilde{Q}}_{s}} \right. \right)\\
s.t.\;&C1:Tr({\mathbf{Q}_s}) \le {P_{\max }}\\
    &C2:{\mathbf{Q}_s} \succcurlyeq 0.
\end{split}
\end{equation}
Then this problem is convex and can be easily solved using standard interior-point methods \cite{DBLP:journals/tvt/CumananDSTL14}. That is, the Karush¨CKuhn¨CTucker (KKT) conditions \cite{DBLP:books/cu/BV2014} for the above convex approximation problem are, namely,
%KKTÌõ¼þ
\begin{equation}\label{eq13}
\left\{ \begin{matrix}
   \frac{1}{ln2}\frac{1}{\sigma _{R}^{2}}\mathbf{G}_{TR}^{H}\mathbf{W}_{R}^{-1}{{\mathbf{G}}_{TR}}&-\frac{1}{ln2}\frac{1}{\sigma _{E}^{2}}\mathbf{G}_{TE}^{H}\mathbf{W}^{-1}_{E,{\mathbf{\tilde{Q}}_s}}{{\mathbf{G}}_{TE}}-\lambda {{\mathbf{I}}_{{{N}_{T}}}}+{{\mathbf{Z}}^{T}}=0  \\
   \lambda \left[ Tr\left( {{\mathbf{Q}}_{s}} \right)-{{P}_{\max }} \right]&=0\text{                                                   }.  \\
\end{matrix} \right.
\end{equation}
Herein, ${\mathbf{W}_R}={\mathbf{I}_{N_R}}+\frac{1}{\sigma _R^2}{\mathbf{G}_{TR}}{\mathbf{Q}_s}\mathbf{G}_{TR}^{H}$. $\lambda \ge 0$ and $\mathbf{Z}\succcurlyeq 0$ are the dual variables associated with the transmission power constraint and the positive semi-definite constraint on the $\mathbf{Q}_s$, respectively. Correspondingly, the Lagrangian function of (\ref{eq12}) can be written as
%À­¸ñÀÊÈÕº¯Êý
\begin{equation}\label{eq14}
L\left( {{\mathbf{Q}}_{s}},\lambda ,\mathbf{Z} \right)=-{{\tilde{R}}_{sec }}+\lambda \left[ Tr\left( {{\mathbf{Q}}_{s}} \right)-{{P}_{\max }} \right]-Tr\left[ \mathbf{Z}{{\mathbf{Q}}_{s}} \right]
\end{equation}
Since problem (\ref{eq12}) is convex and satisfies the Slater's condition, the duality gap is zero between (\ref{eq12}) and its dual problem. Thus, the optimal solution of (\ref{eq12}) can be determined via solving the following Lagrange dual problem
%À­¸ñÀÊÈÕ¶ÔżÎÊÌâ
\begin{equation}\label{eq15}
\underset{\lambda \ge 0}{\mathop{\min }}\,\text{  g}\left( \lambda  \right).
\end{equation}
Herein,
%¶ÔżÎÊÌâ±í´ïʽ
\begin{equation}\label{eq16}
\text{g}\left( \lambda  \right)=\underset{{{\mathbf{Q}}_{s}}\succcurlyeq 0}{\mathop{\max }}\,\left\{ {{{\tilde{R}}}_{sec }}-\lambda \left[ Tr\left( {{\mathbf{Q}}_{s}} \right)-{{P}_{\max }} \right] \right\}.
\end{equation}
To sum up, we have the SCA based suboptimal algorithm for OP2 which is summarized as the Algorithm 1 as below.
%Ëã·¨1£ºÓÅ»¯·¢ÉäЭ·½²î¾ØÕó
\begin{table}[h]
\centering
\begin{tabular}{l}
\hline
\textbf{Algorithm 1: }Optimize transmit covariance matrix\\
\hline
S1: Initialize: $\mathbf{\tilde{Q}}_s\succcurlyeq 0$ and $\lambda = \lambda_0 >0$;\\
S2: Repeat\\
S3: \quad Repeat\\
\qquad\qquad a) Solve the problem in (\ref{eq16}) with given $\mathbf{\tilde{Q}}_s$ and $\lambda$, \\
\qquad\qquad Obtain the optimal transmit covariance ${\mathbf{\hat{Q}}}_s$;\\
\qquad\qquad b) Update $\lambda$ based on the subgradient method;\\
S4: \quad Until the required accuracy;\\
S5: \quad Update $\mathbf{\tilde{Q}}_s=\mathbf{\hat{Q}}_s$, and reset $\lambda = \lambda_0$;\\
S6: Until the required accuracy;\\
S7: Output ${\mathbf{\hat{Q}}}_s$.\\
\hline
\end{tabular}
\label{tab1}
\end{table}

\noindent To meet the transmission power constraint at the AP and the positive semi-definite constraint of the transmit covariance matrix $\mathbf{Q}_s$ at the beginning of the algorithm, we set $\mathbf{\tilde{Q}}_s=\left( P_{\max }/N_T\ \right){{\mathbf{I}}_{{{N}_{T}}}}$.

\subsection{Optimize the IRS reflecting coefficients}
In this subsection, given the transmit covariance matrix $\mathbf{Q}_s$ at the AP, the optimization the reflecting coefficient matrix $\mathbf{\Theta}$ at IRS with the continuous model is discussed. Particularly, we have the following problem OP3.
\begin{small}
\begin{equation}\label{eq17}
\begin{split}
\underset{\mathbf{\Theta} }{\mathop{\max }}\,{{R}_{sec }}\left( \mathbf{\Theta}  \right)&={{\log }_{2}}\det \left( {{\mathbf{I}}_{{{N}_{R}}}}+\frac{1}{\sigma _{R}^{2}}\mathbf{G}_{TR}\left(\mathbf{\Theta}\right){{\mathbf{Q}}_{s}}\mathbf{G}_{TR}^H\left(\mathbf{\Theta}\right) \right)\\
&-{{\log }_{2}}\det \left( {{\mathbf{I}}_{{{N}_{E}}}}+\frac{1}{\sigma _E^2}\mathbf{G}_{TE}\left(\mathbf{\Theta}\right){{\mathbf{Q}}_{s}}\mathbf{G}_{TE}^H\left(\mathbf{\Theta}\right) \right)\\
s.t. &{{\theta }_{m}}\in {\boldsymbol{\Phi }_{1}},m=1,2,...,M\\
\end{split}
\end{equation}
\end{small}
It is obvious that OP3 is a non-convex programming with both non-convex constraints and non-concave objective function, which makes it is difficult to be solved. However, we prove that, given $\left\{ {{\theta }_{i}} \right\}_{i=1,i\ne m}^{M}$, the formulated optimization problem with respect to $\theta _m$ can be solved with the close-form optimal solution or to have an interval about the optimal solution. Therefore, the alternative optimization approach is used here again to solve OP3, i.e., we alternatively solve OP3 in variable $\theta_m$ with given $\theta_i,i=1,...,M, i\neq m$ until the procedure is converged. The details are illustrated as follows.

\emph{1) Objective function transformation:} In order to use the alternative optimization approach to solve OP3, we should first make an objective function transformation for OP3. Note that, the relationship of the objective function with $\left\{{\theta _m} \right\}_{m=1}^M$ is rather implicit. Thus, we rewrite the objective function as an explicit function over $\theta _m , \forall m$. That is \cite{DBLP:journals/corr/abs-1910-01573},
%°²È«ËÙÂÊÏÔʽ±í´ïʽ
\begin{small}
\begin{equation}\label{eq18}
\begin{split}
{R_{sec }}\left(\mathbf{\Theta} \right)&={{\log }_{2}}\det \left( {{\mathbf{I}}_{{{N}_{R}}}}+\frac{1}{\sigma _{R}^{2}}\overline{\mathbf{H}}_{TR}\overline{\mathbf{H}}_{TR}^H \right.+\frac{1}{\sigma _{R}^{2}}\sum\limits_{i=1}^{M}{{{\boldsymbol{h}}_{SR,i}} \overline{\boldsymbol{h}}_{TS,i}^H\overline{\boldsymbol{h}}_{TS,i}\boldsymbol{h}_{SR,i}^{H}}\\
&+\frac{1}{\sigma _{R}^{2}}\sum\limits_{i=1}^{M}{\sum\limits_{j=1,j\ne i}^{M}{{{\theta }_{i}}\theta _{j}^*\boldsymbol{h}_{SR,i}\overline{\boldsymbol{h}}_{TS,i}^H\overline{\boldsymbol{h}}_{TS,j}\boldsymbol{h}_{SR,j}^{H}}}\left. +\frac{1}{\sigma _{R}^{2}}\sum\limits_{i=1}^{M}{\left( \overline{\mathbf{H}}_{TR}\theta _{i}^{*}\overline{\boldsymbol{h}}_{TS,i}\boldsymbol{h}_{SR,i}^{H}+{{\theta }_{i}}{{\boldsymbol{h}}_{SR,i}}\overline{\boldsymbol{h}}_{TS,i}^H\overline{\mathbf{H}}_{TR}^H \right)} \right)\\
&-{{\log }_{2}}\det \left( {\mathbf{I}_{{{N}_{E}}}}+\frac{1}{\sigma _{E}^{2}}\overline{\mathbf{H}}_{TE}\overline{\mathbf{H}}_{TE}^H \right.+\frac{1}{\sigma _{E}^{2}}\sum\limits_{i=1}^{M}{{{\boldsymbol{h}}_{SE,i}}\overline{\boldsymbol{h}}_{TS,i}^H\overline{\boldsymbol{h}}_{TS,i}\boldsymbol{h}_{SE,i}^{H}}\\
&+\frac{1}{\sigma _{E}^{2}}\sum\limits_{i=1}^{M}{\sum\limits_{j=1,j\ne i}^{M}{{{\theta }_{i}}\theta _{j}^{*}{{\boldsymbol{h}}_{SE,i}}\overline{\boldsymbol{h}}_{TS,i}^H\overline{\boldsymbol{h}}_{TS,j}\boldsymbol{h}_{SE,j}^{H}}}\left. +\frac{1}{\sigma _{E}^{2}}\sum\limits_{i=1}^{M}{\left( \overline{\mathbf{H}}_{TE}\theta _{i}^{*}\overline{\boldsymbol{h}}_{TS,i}\boldsymbol{h}_{SE,i}^{H}+{{\theta }_{i}}{{\boldsymbol{h}}_{SE,i}}\overline{\boldsymbol{h}}_{TS,i}^H\overline{\mathbf{H}}_{TE}^H \right)} \right)\\
\end{split}
\end{equation}
\end{small}
Herein, let $\mathbf{Q}_s={\mathbf{U}_Q}{\mathbf{\Sigma} _Q}{\mathbf{U}_Q^H}$ as the eigenvalue decomposition (EVD) of ${\mathbf{Q}_s}\succcurlyeq 0$, ${\mathbf{U}_Q}\in {{\mathbb{C}}^{{N_T}\times {N_T}}}$ and ${\mathbf{\Sigma} _Q}\in {{\mathbb{C}}^{{N_T}\times {N_T}}}$ are unitary matrix and diagonal matrix, respectively, and all the diagonal elements in ${\mathbf{\Sigma} _Q}$ are non-negative real numbers. Also, in (\ref{eq18}), we define ${{\mathbf{H}}_{SR}}=[{{\boldsymbol{h}}_{SR,1}},...,{{\boldsymbol{h}}_{SR,M}}]$, ${{\mathbf{H}}_{SE}}=[{{\boldsymbol{h}}_{SE,1}},...,{{\boldsymbol{h}}_{SE,M}}]$ , $\overline{\mathbf{H}}_{TS}=\mathbf{H}_{TS}\mathbf{U}_Q\mathbf{\Sigma} _{Q}^{{1}/{2}\;}={{[\overline{\boldsymbol{h}}_{TS,1},...,\overline{\boldsymbol{h}}_{TR,M}]}^{H}}$, $\overline{\mathbf{H}}_{TR}=\mathbf{H}_{TR}\mathbf{U}_Q\mathbf{\Sigma} _{Q}^{{1}/{2}\;}\in {{\mathbb{C}}^{{{N}_{R}}\times {{N}_{T}}}}$, $\overline{\mathbf{H}}_{TE}={{\mathbf{H}}_{TE}}{{\mathbf{U}}_{Q}}\mathbf{\Sigma} _{Q}^{{1}/{2}\;}\in {{\mathbb{C}}^{{{N}_{E}}\times {{N}_{T}}}}$, and ${{\boldsymbol{h}}_{SR,m}}\in {{\mathbb{C}}^{{{N}_{R}}\times 1}}$, ${{\boldsymbol{h}}_{SE,m}}\in {{\mathbb{C}}^{{{N}_{E}}\times 1}}$, $\overline{\boldsymbol{h}}_{TS,m}\in {{\mathbb{C}}^{{{N}_{T}}\times 1}},m=1,2,...M$. Now, $R_{sec}$ is represented in an explicit form of the reflection coefficients $\left\{{\theta _m} \right\}_{m=1}^M$. Therefore, given $\mathbf{Q}_s$ and $\left\{ {\theta _i} \right\}_{i=1,i\ne m}^M$, $R_{sec}$ can be rewritten as a function of $\theta _m$ as,
\begin{equation}\label{eq19}
\begin{split}
{{R}_{sec }}\left( {{\theta }_{m}} \right)&={{\log }_{2}}\det \left( {{\mathbf{A}}_{R,m}}+{{\theta }_{m}}{{\mathbf{B}}_{R,m}}+\theta _{m}^{*}\mathbf{B}_{R,m}^{H} \right)-{{\log }_{2}}\det \left( {{\mathbf{A}}_{E,m}}+{{\theta }_{m}}{{\mathbf{B}}_{E,m}}+\theta _{m}^{*}\mathbf{B}_{E,m}^{H} \right),\forall m,\\
\end{split}
\end{equation}
where,
%Çó¹Ø¼ü¾ØÕó
\begin{equation}\label{eq20}
\begin{split}
&{{\mathbf{A}}_{R,m}}={{\mathbf{I}}_{{{N}_{R}}}}+\frac{1}{\sigma _{R}^{2}}\hat{\mathbf{H}}_R\hat{\mathbf{H}}_R^H+\frac{1}{\sigma _{R}^{2}}\tilde{\mathbf{H}}_R\tilde{\mathbf{H}}_R^H,\forall m,\\
&{{\mathbf{B}}_{R,m}}=\frac{1}{\sigma _{R}^{2}}\tilde{\mathbf{H}}_R\hat{\mathbf{H}}_R^{H},\forall m,\\
&{{\mathbf{A}}_{E,m}}={{\mathbf{I}}_{{{N}_{E}}}}+\frac{1}{\sigma _{E}^{2}}\hat{\mathbf{H}}_E\hat{\mathbf{H}}_E^H+\frac{1}{\sigma _{E}^{2}}\tilde{\mathbf{H}}_E\tilde{\mathbf{H}}_E^H,\forall m,\\
&{{\mathbf{B}}_{E,m}}=\frac{1}{\sigma _{E}^{2}}\tilde{\mathbf{H}}_E\hat{\mathbf{H}}_E^H,\forall m.
\end{split}
\end{equation}
We denote $\hat{\mathbf{H}}_R=\overline{\mathbf{H}}_{TR}+\sum\limits_{i=1,i\ne m}^{M}{{{\theta }_{i}}{{\boldsymbol{h}}_{SR,i}}\overline{\boldsymbol{h}}_{TS,i}^H}$, $\hat{\mathbf{H}}_E=\overline{\mathbf{H}}_{TE}+\sum\limits_{i=1,i\ne m}^{M}{{{\theta }_{i}}{{\boldsymbol{h}}_{SE,i}}\overline{\boldsymbol{h}}_{TS,i}^H}$, $\tilde{\mathbf{H}}_R={{\boldsymbol{h}}_{SR,m}}\overline{\boldsymbol{h}}_{TS,m}^H$ and $\tilde{\mathbf{H}}_E={{\boldsymbol{h}}_{SE,m}}\overline{\boldsymbol{h}}_{TS,m}^H$. Since both ${\mathbf{A}}_{R,m}$ and ${\mathbf{A}}_{E,m}$ are the sum of identity matrix and the two positive semi-define matrixes, thus we have $\mathbf{A}_{R,m}\succ 0$, $\mathbf{A}_{E,m}\succ 0$, $rank\left( {{\mathbf{A}}_{R,m}} \right)={{N}_{R}}$ and $rank\left( {{\mathbf{A}}_{E,m}} \right)={{N}_{E}}$. Moreover, for ${\mathbf{B}}_{R,m}$ and ${\mathbf{B}}_{E,m}$ we have $rank\left( {{\mathbf{B}}_{R,m}} \right) \le \\ rank\left( {{\boldsymbol{h}}_{SR,m}}\overline{\boldsymbol{h}}_{TS,m}^H \right)\le 1$ and $rank\left( {{\mathbf{B}}_{E,m}} \right)\le rank\left( {{\boldsymbol{h}}_{SE,m}}\overline{\boldsymbol{h}}_{TS,m}^H \right)\le 1$, respectively. Therefore, ${{R}_{sec }}\left( {{\theta }_{m}} \right)$ can be rewritten as
\begin{equation}\label{eq21}
\begin{split}
{{R}_{sec }}\left( {{\theta }_{m}} \right)&={{\log }_{2}}\det \left( {{\mathbf{I}}_{{{N}_{R}}}}+{{\theta }_{m}}\mathbf{J}_R+\theta _{m}^{*}\mathbf{J}_R^{H} \right)-{{\log }_{2}}\det \left( {{\mathbf{I}}_{{{N}_{E}}}}+{{\theta }_{m}}\mathbf{J}_E+\theta _{m}^{*}\mathbf{J}_E^{H} \right)\\
&+{{\log }_{2}}\det \left( {{\mathbf{A}}_{R,m}} \right)-{{\log }_{2}}\det \left( {{\mathbf{A}}_{E,m}} \right)\\
&=\overline{R}_{sec }\left( {{\theta }_{m}} \right)+{{\log }_{2}}\det \left( {{\mathbf{A}}_{R,m}} \right)-{{\log }_{2}}\det \left( {{\mathbf{A}}_{E,m}} \right)\\
\end{split}
\end{equation}
Herein, $\mathbf{J}_R=\mathbf{A}_{R,m}^{-1}{{\mathbf{B}}_{R,m}}$ and $\mathbf{J}_E=\mathbf{A}_{E,m}^{-1}{{\mathbf{B}}_{E,m}}$. Hence, the maximization of ${R}_{sec }$ is equivalent to maximize the $\overline{R}_{sec }$, namely,
\begin{equation}\label{eq22}
\begin{split}
\overline{R}_{sec }\left( {{\theta }_{m}} \right)&={{\log }_{2}}\det \left( {{\mathbf{I}}_{{{N}_{R}}}}+{{\theta }_{m}}\mathbf{J}_R+\theta _{m}^{*}\mathbf{J}_R^H \right)-{{\log }_{2}}\det \left( {{\mathbf{I}}_{{{N}_{E}}}}+{{\theta }_{m}}\mathbf{J}_E+\theta _{m}^{*}\mathbf{J}_E^H \right)\\
&= \overline{R}_R(\theta_m)-\overline{R}_{E}(\theta_m)\\
s.t.&\left| {{\theta }_{m}} \right|\text{=1 }\\
\end{split}
\end{equation}
Herein, $\overline{R}_R(\theta_m)=\log_2\det(\mathbf{I}_{N_R}+\theta_m \mathbf{J}_R+\theta^{*}_{m}\mathbf{J}^{H}_{R})$ and $\overline{R}_E(\theta_m)=\log_2\det(\mathbf{I}_{N_E}+\theta_m\mathbf{J}_E+\theta^{*}_{m}\mathbf{J}^{H}_{E})$. In addition, due to both $\mathbf{A}_{R,m}$ and $\mathbf{A}_{E,m}$ are full-rank, we have $rank\left( \mathbf{J}_R \right)=rank\left( \mathbf{B}_{R,m} \right)\le 1$ and $rank\left( \mathbf{J}_E \right)=rank\left( \mathbf{B}_{E,m} \right)\le 1$.

\emph{2) Deriving the tractable expressions for $\overline{R}_{R}(\theta_m)$ and $\overline{R}_E(\theta_m)$\cite{DBLP:journals/corr/abs-1910-01573}:} Following the above, herein, according to the value of $rank\left( \mathbf{J}_R \right)$ (or $rank\left( \mathbf{J}_E \right)$), i.e., $rank\left( \mathbf{J}_R \right)=1$ ($rank\left( \mathbf{J}_E \right)=1$) or $rank\left( \mathbf{J}_R \right)=0$ ($rank\left( \mathbf{J}_E \right)=0$), we separately derive the tractable expressions of $\overline{R}_R(\theta_m)$ and $\overline{R}_{E}(\theta_m)$ which are then used to analyze the corresponding optimal solution of $\theta_m$.

\underline{\emph{Case $rank\left( \mathbf{J}_R \right)=1$}}: At first, we present a lemma as below.

\textbf{Lemma 1} (\cite{DBLP:journals/corr/abs-1910-01573}): $\mathbf{J}_R$ is diagonalizable if and only if $Tr\left( \mathbf{J}_R \right) \ne 0$ .

Based on the Lemma 1, then we can derive the expression of $\overline{R}_R\left( {{\theta }_{m}} \right)$ under $Tr\left( \mathbf{J}_R \right)= 0$ and $Tr\left( \mathbf{J}_R \right) \ne 0$, separately.

\textbf{If $Tr\left( \mathbf{J}_R\right)=0$}, namely, $\mathbf{J}_R$ is non-diagonalizable, $\mathbf{J}_R={{\boldsymbol{u}}_{R,m}}\boldsymbol{v}_{R,m}^{H}$ with ${{\boldsymbol{u}}_{R,m}},{{\boldsymbol{v}}_{R,m}}\in {{\mathbb{C}}^{{{N}_{R}}\times 1}}$ and $\boldsymbol{v}_{R,m}^{H}{{\boldsymbol{u}}_{R,m}}=\boldsymbol{u}_{R,m}^{H}{{\boldsymbol{v}}_{R,m}}=Tr\left( \mathbf{J}_R \right)=0$ due to $rank\left( \mathbf{J}_R \right)=1$. Hence, the expression of $\overline{R}_R\left( {{\theta }_{m}} \right)$ can be transformed into
\begin{equation}\label{eq23}
\begin{split}
\overline{R}_R\left( {{\theta }_{m}} \right)&={{\log }_{2}}\det \left( {{\mathbf{I}}_{{{N}_{R}}}}-\mathbf{A}_{R,m}^{-1}{{\boldsymbol{v}}_{R,m}}\boldsymbol{u}_{R,m}^{H}{{\mathbf{A}}_{R,m}}{{\boldsymbol{u}}_{R,m}} \notag\boldsymbol{v}_{R,m}^{H} \right)\\
&={{\log }_{2}}\det \left( {{\mathbf{I}}_{{{N}_{R}}}}-\mathbf{A}_{R,m}^{-1}\mathbf{J}_{R}^{H}{{\mathbf{A}}_{R,m}}{{\mathbf{J}}_{R}} \right)\\
&={{\log }_{2}}\det \left( {{\mathbf{I}}_{{{N}_{R}}}}-\mathbf{A}_{R,m}^{-1}\mathbf{B}_{R,m}^{H}\mathbf{J}_R \right)\\
\end{split},
\end{equation}
in which, the last equation is hold with $\mathbf{A}_{R,m}^{-1} = \left(\mathbf{A}_{R,m}^{-1}\right)^H$  for $\mathbf{A}_{R,m}\succ 0$.

\textbf{If $Tr\left( \mathbf{J}_R \right) \ne 0$}, the EVD of $\mathbf{J}_R$ can be expressed as $\mathbf{J}_R={{\mathbf{U}}_{R,m}}{{\mathbf{\Sigma} }_{R,m}}\mathbf{U}_{R,m}^{-1}$, where, ${{\mathbf{U}}_{R,m}}\in {{\mathbb{C}}^{{{N}_{R}}\times {{N}_{R}}}}$ and ${{\mathbf{\Sigma} }_{R,m}}=diag\left\{ {{\lambda }_{R,m}},0,...,0 \right\}$ with $\lambda _{R,m}$ denoting the sole non-zero eigenvalue of $\mathbf{J}_R$. Set ${{\mathbf{V}}_{R,m}}=\mathbf{U}_{R,m}^{H}{{\mathbf{A}}_{R,m}}{{\mathbf{U}}_{R,m}}$ and it is a Hermitian matrix with $\mathbf{V}_{R,m}=\mathbf{V}_{R,m}^H$. Let ${{\boldsymbol{v}}_{R,m}}\in {{\mathbb{C}}^{{{N}_{R}}\times 1}}$ and $\overline{\boldsymbol{v}}_{R,m}^T\in {{\mathbb{C}}^{1\times {{N}_{R}}}}$ denote the first column of $\mathbf{V}_{R,m}^{-1}$ and the first row of $\mathbf{V}_{R,m}$. Note that it follows that $\overline{\boldsymbol{v}}_{R,m}^T{{\boldsymbol{v}}_{R,m}}=1$; moreover, let $v_{R,m1}$ and $\overline{v}_{R,m1}$ denote the first element in $\boldsymbol{v}_{R,m}$ and $\overline{\boldsymbol{v}}_{R,m}^T$, respectively, we have $\overline{v}_{R,m1}v_{R,m1} \in \mathbb{R}$ since both $\mathbf{V}_{R,m}$ and $\mathbf{V}_{R,m}^{-1}$ are Hermitian matrices. Hence, $\overline{R}_R\left( {{\theta }_{m}} \right)$ can be further simplified as \cite{DBLP:journals/corr/abs-1910-01573},
\begin{equation}\label{eq24}
\begin{split}
\overline{R}_R\left( {{\theta }_{m}} \right)&={{\log }_{2}}\left( 1+{{\left| {{\lambda }_{R,m}} \right|}^{2}}\left( 1-\overline{v}_{R,m1}{{v}_{R,m1}} \right)\right.+2\Re \left\{ {{\theta }_{m}}{{\lambda }_{R,m}} \right\} \Big)\\
\end{split}
\end{equation}

\underline{\emph{Case $rank\left( \mathbf{J}_E\right)=1$}}: Similarly, \textbf{if $Tr\left( \mathbf{J}_E \right)=0$}, we have
\begin{equation}\label{eq25}
\overline{R}_E\left( {{\theta }_{m}} \right)={{\log }_{2}}\det \left( {{\mathbf{I}}_{{{N}_{E}}}}-\mathbf{A}_{E,m}^{-1}\mathbf{B}_{E,m}^{H}\mathbf{J}_E \right)
\end{equation}

And \textbf{if $Tr\left( \mathbf{J}_E \right) \ne 0$}, the EVD of $\mathbf{J}_E$ can be expressed as $\mathbf{J}_E={{\mathbf{U}}_{E,m}}{{\mathbf{\Sigma} }_{E,m}}\mathbf{U}_{E,m}^{-1}$, where, ${{\mathbf{U}}_{E,m}}\in {{\mathbb{C}}^{{{N}_{E}}\times {{N}_{E}}}}$ and ${{\mathbf{\Sigma} }_{E,m}}=diag\left\{ {{\lambda }_{E,m}},0,...,0 \right\}$ with $\lambda _{E,m}$ denoting the sole non-zero eigenvalue of $\mathbf{J}_E$. Set ${{\mathbf{V}}_{E,m}}=\mathbf{U}_{E,m}^{H}{{\mathbf{A}}_{E,m}}{{\mathbf{U}}_{E,m}}$, let ${{\boldsymbol{v}}_{E,m}}\in {{\mathbb{C}}^{{{N}_{E}}\times 1}}$ and $\overline{\boldsymbol{v}}_{E,m}^T\in {{\mathbb{C}}^{1\times {{N}_{E}}}}$ denote the first column of $\mathbf{V}_{E,m}^{-1}$ and the first row of $\mathbf{V}_{E,m}$ and let $v_{E,m1}$ and $\overline{v}_{E,m1}$ denote the first element in $\boldsymbol{v}_{E,m}$ and $\overline{\boldsymbol{v}}_{E,m}^T$, respectively. Hence, $\overline{R}_E\left( {{\theta }_{m}} \right)$ can be further simplified as,
\begin{equation}\label{eq26}
\begin{split}
\overline{R}_E\left( {{\theta }_{m}} \right)&={{\log }_{2}}\left( 1+{{\left| {{\lambda }_{E,m}} \right|}^{2}}\left( 1-\overline{v}_{E,m1}{{v}_{E,m1}} \right)\right.+2\Re \left\{ {{\theta }_{m}}{{\lambda }_{E,m}} \right\} \Big)\\
\end{split}
\end{equation}

\underline{\emph{Case $rank(\mathbf{J}_R)=0$ or $rank(\mathbf{J}_E)=0$}}: In fact, if $rank( \mathbf{J}_R )$ $ = 0$, we always have $\mathbf{J}_R=\mathbf{0}$ and $\mathbf{J}_R={{\boldsymbol{u}}_{R,m}}\boldsymbol{v}_{R,m}^{H}$ with ${{\boldsymbol{u}}_{R,m}}={{\boldsymbol{v}}_{R,m}}=\mathrm{0}$, which is equivalent to the case $Tr\left( \mathbf{J}_R \right)=0$ under $rank\left( \mathbf{J}_R \right)=1$. Similarly, the case $rank\left( \mathbf{J}_E \right)=0$ is equivalent to the case $Tr\left(\mathbf{J}_E \right)=0$ under $rank\left( \mathbf{J}_E \right)=1$. Therefore, no matter whether $rank(\mathbf{J}_R)=1$ or $rank(\mathbf{J}_R)=0$ ($rank(\mathbf{J}_E)=1$ or $rank(\mathbf{J}_E)=0$), a tractable expression of $\overline{R}_R(\theta_m)$ ($\overline{R}_E(\theta_m)$) only depends on the value of $Tr(\mathbf{J}_R)$ ($Tr(\mathbf{J}_E)$), i.e., $Tr(\mathbf{J}_R) =0$ or $Tr(\mathbf{J}_R)\ne 0$ ($Tr(\mathbf{J}_E) =0$ or $Tr(\mathbf{J}_E)\ne 0$).

\emph{3) Solving problem (\ref{eq22}):} Based on the deriving of the tractable expressions for $\overline{R}_R(\theta_m)$ and $\overline{R}_E(\theta_m)$, we know that problem (\ref{eq22}) should be discussed and solved by considering four different conditions, i.e., $Tr(\mathbf{J}_R)=0$ and $Tr(\mathbf{J}_E)=0$, $Tr(\mathbf{J}_R)\ne 0$ and $Tr(\mathbf{J}_E)=0$, $Tr(\mathbf{J}_R)=0$ and $Tr(\mathbf{J}_E)\ne 0$, and $Tr(\mathbf{J}_R)\ne 0$ and $Tr(\mathbf{J}_E) \ne 0$, as follows.

\underline{\emph{Case $Tr\left(\mathbf{J}_R\right)=Tr\left(\mathbf{J}_E\right) = 0$:}} Namely, both $\mathbf{J}_R$ and $\mathbf{J}_E$ are non-diagonalizable. In this case, $\overline{R}_{sec }\left( {{\theta }_{m}} \right)$ is defined as
\begin{equation}\label{eq27}
\begin{split}
\overline{R}_{sec }\left( {{\theta }_{m}} \right)&={{\log }_{2}}\det \left( {{\mathbf{I}}_{{{N}_{R}}}}-\mathbf{A}_{R,m}^{-1}{\mathbf{B}_{R,m}^H}\mathbf{J}_R \right)-{{\log }_{2}}\det \left( {{\mathbf{I}}_{{{N}_{E}}}}-\mathbf{A}_{E,m}^{-1}{\mathbf{B}_{E,m}^H}\mathbf{J}_E \right).\\
\end{split}
\end{equation}
That is, $\overline{R}_{sec }\left( {{\theta }_{m}} \right)$ is independent of $\theta _m$. Hence, we can directly obtain the optimal solution for (\ref{eq22}) and it is characterized by the following proposition. Since the proof is simply thus it is omitted here for simplification.

\textbf{Proposition 2:} If $Tr\left(\mathbf{J}_R\right)=Tr\left(\mathbf{J}_E\right) = 0$, any $\theta _m$ with $\left| {\theta _m} \right|=1$ is the optimal solution for (\ref{eq22}) and the corresponding optimal value is
\begin{equation}\label{eq28}
\begin{split}
{{\hat{R}}_{sec }}\left( {{\theta }_{m}} \right)&={{\log }_{2}}\det \left( {{\mathbf{A}}_{R,m}}-\mathbf{B}_{R,m}^{H}\mathbf{J}_R \right)-{{\log }_{2}}\det \left( {{\mathbf{A}}_{E,m}}-\mathbf{B}_{E,m}^{H}\mathbf{J}_E \right).\\
\end{split}
\end{equation}

\underline{\emph{Case $Tr\left(\mathbf{J}_R\right) \ne 0$ and $Tr\left(\mathbf{J}_E\right) = 0$:}} Namely, $\mathbf{J}_R$ is diagonalizable and $\mathbf{J}_E$ is non-diagonalizable. In this case, $\overline{R}_{sec }\left( {{\theta }_{m}} \right)$ is denoted as
\begin{equation}\label{eq29}
\begin{split}
\overline{R}_{sec }\left( {{\theta }_{m}} \right)&={{\log }_{2}}\left( 1+{{\left| {{\lambda }_{R,m}} \right|}^{2}}\left( 1-\overline{v}_{R,m1}{{v}_{R,m1}} \right)\right.+2\Re \left\{ {{\theta }_{m}}{{\lambda }_{R,m}} \right\} \Big)-{{\log }_{2}}\det \left( {{\mathbf{I}}_{{{N}_{E}}}}-\mathbf{A}_{E,m}^{-1}{\mathbf{B}_{E,m}^H}\mathbf{J}_E \right).\\
\end{split}
\end{equation}
Now, for problem (\ref{eq22}), maximizing $\overline{R}_{sec}(\theta_m)$ is equivalent to maximize $\Re \left\{ {{\theta }_{m}}{{\lambda }_{E,m}} \right\}$ in (\ref{eq29}) and the corresponding optimal solution can be characterized by the following Proposition 3.

\textbf{Proposition 3:} If $Tr\left(\mathbf{J}_R\right) \ne 0$ and $Tr\left(\mathbf{J}_E\right) = 0$, the optimal solution to (\ref{eq22}) is
\begin{equation*}
{{\hat{\theta }}_{m}}={{e}^{-j\arg \left( {{\lambda }_{R,m}} \right)}},
\end{equation*}
and the corresponding optimal value is
\begin{equation*}
\begin{split}
{{{\hat{R}}}_{sec }}\left( {{\theta }_{m}} \right)&={{\log }_{2}}\left( 1+{{\left| {{\lambda }_{R,m}} \right|}^{2}}\left( 1-\overline{v}_{R,m1}{{v}_{R,m1}} \right)\right.+2\left| {{\lambda }_{R,m}} \right| \Big)+{{\log }_{2}}\det \left( {{\mathbf{A}}_{R,m}} \right)\\
&-{{\log }_{2}}\det \left( {{\mathbf{A}}_{E,m}}-\mathbf{B}_{E,m}^{H}\mathbf{J}_E \right).\\
\end{split}
\end{equation*}

\emph{Proof:} The proof is provided in Appendix \ref{appendix_1}. $\hfill\blacksquare$
%\emph{Proof:} Please refer to Appendix \ref{appendix_1}. $\hfill\blacksquare$

\underline{\emph{Case $Tr\left(\mathbf{J}_R\right)=0$ and $Tr\left(\mathbf{J}_E\right) \ne 0$:}} Namely, $\mathbf{J}_R$ is non-diagonalizable and $\mathbf{J}_E$ is diagonalizable, then $\overline{R}_{sec }\left( {{\theta }_{m}} \right)$ becomes
\begin{equation}\label{eq30}
\begin{split}
\overline{R}_{sec }\left( {{\theta }_{m}} \right)&={{\log }_{2}}\det \left( {{\mathbf{I}}_{{{N}_{R}}}}-\mathbf{A}_{R,m}^{-1}{\mathbf{B}_{R,m}^H}\mathbf{J}_R \right)-{{\log }_{2}}\left( 1+{{\left| {{\lambda }_{E,m}} \right|}^{2}}\left( 1-\overline{v}_{E,m1}{{v}_{E,m1}} \right)\right.+2\Re \left\{ {{\theta }_{m}}{{\lambda }_{E,m}} \right\} \Big).\\
\end{split}
\end{equation}
Now, for problem (\ref{eq22}), maximizing $\overline{R}_{sec}(\theta_m)$ is equivalent to minimize $\Re \left\{ {{\theta }_{m}}{{\lambda }_{E,m}} \right\}$ in (\ref{eq30}) and the corresponding optimal solution can be characterized by the following Proposition 4.

\textbf{Proposition 4}: If $Tr\left(\mathbf{J}_R\right)=0$ and $Tr\left(\mathbf{J}_E\right) \ne 0$, the optimal solution for (\ref{eq22}) is
\begin{equation*}
{{\hat{\theta }}_{m}}={{e}^{j\left( \pi -\arg \left( {{\lambda }_{E,m}} \right) \right)}},
\end{equation*}
and the corresponding optimal value is
\begin{equation*}
\begin{split}
{{{\hat{R}}}_{sec }}\left( {{\theta }_{m}} \right)&={{\log }_{2}}\det \left( {{\mathbf{A}}_{R,m}}-\mathbf{B}_{R,m}^{H}\mathbf{J}_R \right)-{{\log }_{2}}\left( 1+{{\left| {{\lambda }_{E,m}} \right|}^{2}}\left( 1-\overline{v}_{E,m1}{{v}_{E,m1}} \right)\right.-2\left| {{\lambda }_{E,m}} \right| \Big)\\
&-{{\log }_{2}}\det \left( {{\mathbf{A}}_{E,m}} \right).\\
\end{split}
\end{equation*}

\emph{Proof:} The proof is provided in Appendix \ref{appendix_2}. $\hfill\blacksquare$
%\emph{Proof:} Please refer to Appendix \ref{appendix_2}. $\hfill\blacksquare$

\underline{\emph{Case $Tr\left(\mathbf{J}_R\right) \ne 0$ and $Tr\left(\mathbf{J}_E\right) \ne 0$:}} Namely, both $\mathbf{J}_R$ and $\mathbf{J}_E$ are diagonalizable. Before presenting the optimal solution conclusion for this case, we firstly introduce a lemma as below.

\textbf{Lemma 5:} For function $f(x)=(a+b\cos x)/[c+d\cos (x+\omega)]$ with $a>b>0,c>d>0,\omega \in \left[ 0,2\pi  \right)$, and the variable $x\in \left[ 0,2\pi  \right)$, we have,

(i) If $\omega \in \left[ 0,\pi \right)$, for $\forall x\in[0,2\pi)$, $\exists \hat{x}\in [0,\pi-\omega]$ satisfies $f(\hat{x})\geq f(x)$, i.e., for $\forall x\in[0,2\pi)$, there always exists a optimal solution $\hat{x}\in [0,\pi-\omega]$ maximizes $f(x)$;

(ii) If $\omega \in \left[ \pi,2\pi  \right)$, for $\forall x\in[0,2\pi)$, $\exists\hat{x}\in \left[ 3\pi-\omega,2\pi \right]$ satisfies $f\left( {\hat{x}} \right)\ge f\left( x \right),\forall x\in \left[ 0,2\pi  \right)$, i.e., for $\forall x\in[0,2\pi)$, there always exists a optimal solution $\hat{x}\in \left[ 3\pi-\omega,2\pi \right]$\footnote{Herein, the variable $x$ can take the value of $2\pi$ in the optimal interval, however, the definition domain of the $f(x)$ is a right-open interval, i.e., $x\in[0,2\pi)$. In fact, it is not conflict with each other as $f(2\pi)=f(0)$.}  maximizes $f(x)$.

\emph{Proof:} The proof is provided in Appendix \ref{appendix_3}.$\hfill\blacksquare$
%\emph{Proof:} Please refer to Appendix \ref{appendix_3}. $\hfill\blacksquare$

Based on the Lemma 5, let $\alpha _R=1+|\lambda _{R,m}|^2(1-\overline{v}_{R,m1}v_{R,m1})$, $\beta _R=2|\lambda _{R,m}|$, $\alpha _E=1+|\lambda _{E,m}|^2(1-\overline{v}_{E,m1}v_{E,m1})$, $\beta _E=2|\lambda _{E,m}|$, $\varphi _m=\arg (\theta _m)$, $\varphi _{\lambda _{R,m}}=\arg (\lambda _{R,m})$, $\varphi _{\lambda _{E,m}}=\arg (\lambda _{E,m})$, then $\overline{R}_{sec }\left( {{\theta }_{m}} \right)$  is rewritten as
\begin{equation}\label{eq31}
\begin{split}
\overline{R}_{sec }\left( {{\theta }_{m}} \right)&={{\log }_{2}}\left( 1+{{\left| {{\lambda }_{R,m}} \right|}^{2}}\left( 1-\overline{v}_{R,m1}{{v}_{R,m1}} \right)+2\Re \left\{ {{\theta }_{m}}{{\lambda }_{R,m}} \right\} \right)\\
&-{{\log }_{2}}\left( 1+{{\left| {{\lambda }_{E,m}} \right|}^{2}}\left( 1-\overline{v}_{E,m1}{{v}_{E,m1}} \right)+2\Re \left\{ {{\theta }_{m}}{{\lambda }_{E,m}} \right\} \right)\\
&={{\log }_{2}}\left( \frac{{{\alpha }_{R}}+{{\beta }_{R}}\cos \left( {{\varphi }_{m}}+{{\varphi }_{{{\lambda }_{R,m}}}} \right)}{{{\alpha }_{E}}+{{\beta }_{E}}\cos \left( {{\varphi }_{m}}+{{\varphi }_{{{\lambda }_{E,m}}}} \right)} \right)\\
&\buildrel \Delta \over = \log_2(f(\varphi _m))\\
\end{split}.
\end{equation}
Now, for problem (\ref{eq22}), maximizing $\overline{R}_{sec}(\theta_m)$ is equivalent to maximize $f\left( {{\varphi }_{m}} \right)$. Moreover, in (\ref{eq31}), we have $1+{{\left| {{\lambda }_{R,m}}\right|}^{2}}( 1-\overline{v}_{R,m1}{{v}_{R,m1}})+2\Re \left\{ {{\theta }_{m}}{{\lambda }_{R,m}} \right\}>0$ and $1+{{\left| {{\lambda }_{E,m}} \right|}^{2}}\left( 1-\overline{v}_{E,m1}{{v}_{E,m1}} \right)+2\Re \left\{ {{\theta }_{m}}{{\lambda }_{E,m}} \right\}>0$ for $\forall {{\theta }_{m}}$, namely, ${{\alpha }_{R}}>{{\beta }_{R}}>0$ and ${{\alpha }_{E}}>{{\beta }_{E}}>0$. Hence, the corresponding optimal solution can be characterized by the Proposition 6 as below and the proof is omitted for it can be easily proved from the Lemma 5.

\textbf{Proposition 6:} If $Tr\left(\mathbf{J}_R\right) \ne 0$ and $Tr\left(\mathbf{J}_E\right) \ne 0$, the optimal solution for (\ref{eq22}) is over the interval defined as below,

(i) If $\bmod \left( {{\varphi }_{{{\lambda }_{E,m}}}}-{{\varphi }_{{{\lambda }_{R,m}}}},2\pi  \right)\in \left[ 0,\pi  \right)$, the optimal phase ${{\hat{\varphi }}_{m}}\in \left( -{{\varphi }_{{{\lambda }_{R,m}}}},\pi -{{\varphi }_{{{\lambda }_{E,m}}}} \right)$;

(ii) If $\bmod \left( {{\varphi }_{{{\lambda }_{E,m}}}}-{{\varphi }_{{{\lambda }_{R,m}}}},2\pi  \right)\in \left[ \pi,2\pi  \right)$, the optimal phase ${{\hat{\varphi }}_{m}}\in \left( 3\pi \text{-}{{\varphi }_{{{\lambda }_{E,m}}}},2\pi -{{\varphi }_{{{\lambda }_{R,m}}}} \right)$.

\noindent In which, $\bmod \left( x,y \right)$ represents the remainder of variable $x$ over $y$. $\hfill\blacksquare$

Based on the Proposition 6, we can then perform the linear search to obtain the optimal solution for problem (\ref{eq22}) under the condition $Tr\left(\mathbf{J}_R\right) \ne 0$ and $Tr\left(\mathbf{J}_E\right) \ne 0$.

To summarize the above analysis, for problem (\ref{eq22}), its optimal solution can be characterized as below
\begin{equation}\label{eq33}
    {{\hat{\theta }}_{m}}=\left\{
        \begin{split}
           &1,&Tr\left(\mathbf{J}_R\right) = 0,Tr\left(\mathbf{J}_E\right) = 0\\
           &{{e}^{-j\arg \left( {{\lambda }_{R,m}} \right)}},&Tr\left(\mathbf{J}_R\right) \ne 0,Tr\left(\mathbf{J}_E\right) = 0\\
           &{{e}^{j\left( \pi -\arg \left( {{\lambda }_{E,m}} \right)\right)}}, &Tr\left(\mathbf{J}_R\right) = 0,Tr\left(\mathbf{J}_E\right) \ne 0\\
           &\overline{\theta} _{m},&Tr\left(\mathbf{J}_R\right) \ne 0,Tr\left(\mathbf{J}_E\right) \ne 0\\
        \end{split}
    \right. ,
\end{equation}
where $\overline{\theta} _{m}$ is obtained by linear search. Therefore, we formulate the algorithm for OP3 as below, i.e., the Algorithm 2.

%Ëã·¨2£ºÓÅ»¯IRS·´ÉäϵÊý
\begin{table}[H]
    \centering
    \begin{tabular}{l}
        \hline
        \textbf{Algorithm 2:} Optimize IRS reflection coefficients\\
        \hline
        S1: Initialize: Randomly generate $\left\{ \theta _{m}^{0} \right\}_{m=1}^{M}, \varepsilon >0, n=0$;\\
        S2: Obtain $\hat{\theta }_{m}^{n},m=1,2,...,M$ by turns according to (\ref{eq33});\\
        S3: If $\sum\limits_{m=1}^{M}{\left| \hat{\theta }_{m}^{n}-\theta _{m}^{n} \right|}>\varepsilon $, set $\left\{ \theta _{m}^{n+1}=\hat{\theta }_{m}^{n} \right\}_{m=1}^{M}, n=n+1$,\\ \quad\quad go back to S2; else set $\left\{ {{{\hat{\theta }}}_{m}}=\theta _{m}^{n} \right\}_{m=1}^{M}$;\\
        S4: Output $\left\{ {{{\hat{\theta }}}_{m}} \right\}_{m=1}^{M}$;\\
        \hline
    \end{tabular}
    \label{tab2}
\end{table}
\noindent In Algorithm 2, $\left\{ {{{\hat{\theta }}}^{n}_{m}} \right\}_{m=1}^{M}$ denotes the solution obtained in the $n$th iteration. The core idea of the algorithm is that, we alternatively optimize $\theta_m$ for given $\theta_i,i=1,...,M,i\ne m$. In addition, the optimization of $\theta_m$ is following the rule of (\ref{eq33}). Furthermore, since the original problem is bounded and the progress of the alternative optimization is monotonically non-decreasing, thus the above algorithm is surely convergent.

\subsection{Overall Algorithm}
In this subsection, the overall algorithm for OP1 is provided. As mentioned, the algorithm is based on alternating optimization, which optimize the objective function with respect to different subsets of optimization variables in each iteration while the other subsets are fixed. Therefore, it is summarized as the Algorithm 3 as follows.
%Ëã·¨3£ºÕûÌåËã·¨
\begin{table}[H]
    \centering
    \begin{tabular}{l}
        \hline
        \textbf{Algorithm 3:}  Alternating Optimization Based Algorithm\\
        \hline
        S1: Initialize: Randomly generate $\left\{ \theta _{m}^{0} \right\}_{m=1}^{M},\mathbf{\tilde{Q}}_{s}^{0}\succcurlyeq 0,\varepsilon >0,n=0$;\\
        S2: Obtain $\mathbf{\hat{Q}}_s^n$ with given $\left\{ \theta _{m}^{n} \right\}_{m=1}^{M}$ and $\mathbf{\tilde{Q}}_{s}^{n}$ by Algorithm 1;\\
        S3: Obtain $\left\{ \hat{\theta }_{m}^{n} \right\}_{m=1}^{M}$ with $\mathbf{\hat{Q}}_s^n$ by Algorithm 2;\\
        S4: If $\sum\limits_{m=1}^{M}{\left| \hat{\theta }_{m}^{n}-\theta _{m}^{n} \right|}>\varepsilon $, set $\left\{ \theta _{m}^{n+1}=\hat{\theta }_{m}^{n} \right\}_{m=1}^{M},\mathbf{\tilde{Q}}_{s}^{n+1}=\mathbf{\hat{Q}}_{s}^{n}$,\\
         \quad $n=n+1$, go back to S2; else set $\left\{ {{{\hat{\theta }}}_{m}}=\theta _{m}^{n} \right\}_{m=1}^{M}$ and ${{\mathbf{\hat{Q}}}_{s}}=\mathbf{\hat{Q}}_{s}^{n}$;\\
        S5: Output $\left\{ {{{\hat{\theta }}}_{m}} \right\}_{m=1}^{M}$ and ${\mathbf{\hat{Q}}}_s$.\\
        \hline
    \end{tabular}
    \label{tab3}
\end{table}

\noindent where $\mathbf{\hat{Q}}_{s}^{n}$ and $\left\{ \hat{\theta }_{m}^{n} \right\}_{m=1}^{M}$ denote the stationary points obtained by the Algorithm 1 and Algorithm 2 in the $n$th iteration of the Algorithm 3, respectively. The procedures of the algorithm are as follows: firstly, we randomly generate a set of $\left\{ \theta _{m}^{0} \right\}_{m=1}^{M}$ with $\left| \theta _{m}^{0} \right|=1,\forall m$ and the phases of $\theta _{m}^{0}$ are following the uniform distribution over $\left[0, 2\pi\right)$. Secondly, for given the reflecting coefficients at the IRS and based on the Algorithm 1, we optimize the transmit covariance matrix at the AP. Thirdly, with the optimized transmit covariance matrix at the AP and based on the Algorithm 2, we optimize the reflecting coefficients at the IRS. Finally, the above two steps are iteratively performed until it is converged. Apparently, duo to the monotonic non-decreasing properties of the Algorithm 1 and the Algorithm 2, and also the objective function in OP1 is bounded, then the overall algorithm is surly converged.

\subsection{Extended to discrete model}
In the previous discussion, we assume that the phase of IRS reflecting element is continuously adjustable, which is too ideal to achieve in practical. And existing studies showed that \cite{DBLP:journals/cm/LiaskosNTPIA18},\cite{DBLP:journals/access/ChenLPG19}, the IRS only can adjust the phase with limited accuracy due to the hardware limitations. To ensure the practicability of the proposed algorithm in more practical application scenarios, the IRS discrete reflecting coefficient model is discussed briefly. In particular, replacing the $\boldsymbol{\Phi} _1$ with $\boldsymbol{\Phi} _2$ in OP1, which formulates the problem OP4 as below.
%IRS·´ÉäϵÊýÓÅ»¯ÎÊÌâÀ©Õ¹µ½ÀëɢģÐÍ
\begin{small}
\begin{equation}\label{eq_dis_opt}
\begin{split}
\underset{\mathbf{\Theta} }{\mathop{\max }}\,{{R}_{sec }}\left( \mathbf{\Theta}  \right)&={{\log }_{2}}\det \left( {{\mathbf{I}}_{{{N}_{R}}}}+\frac{1}{\sigma _{R}^{2}}\mathbf{G}_{TR}\left(\mathbf{\Theta}\right){{\mathbf{Q}}_{s}}\mathbf{G}_{TR}^H\left(\mathbf{\Theta}\right) \right)\\
&-{{\log }_{2}}\det \left( {{\mathbf{I}}_{{{N}_{E}}}}+\frac{1}{\sigma _E^2}\mathbf{G}_{TE}\left(\mathbf{\Theta}\right){{\mathbf{Q}}_{s}}\mathbf{G}_{TE}^H\left(\mathbf{\Theta}\right) \right)\\
s.t.\,&{{\theta }_{m}}\in {\boldsymbol{\Phi }_{2}},m=1,2,...,M\\
\end{split}
\end{equation}
\end{small}
Note that, OP4 belongs to the mixed integer nonlinear programming (MINLP), which is an NP-hard problem and difficult to handle. Hence, the heuristic projection method is used here to solve this problem \cite{DBLP:journals/access/ChenLPG19}. That is, we denote the solution of IRS reflecting coefficients for OP1 and OP4 are ${{\boldsymbol{v}}_{s}}=\left[ {{\theta }_{s,1}},{{\theta }_{s,2}},...,{{\theta }_{s,M}} \right]$ and ${{\boldsymbol{v}}_{d}}=\left[ {{\theta }_{d,1}},{{\theta }_{d,2}},...,{{\theta }_{d,M}} \right]$, respectively. Then ${{\boldsymbol{v}}_{d}}$ is provided by
%×îÓŽâµÄÀëÉ¢À©Õ¹
\begin{equation} \label{eq_dis}
{{\theta }_{d,m}}={{\theta }_{{\hat{q}}}},\hat{q}=\underset{0\le q<Q}{\mathop{\arg \min }}\,\left| {{\theta }_{s,m}}-{{e}^{j{{\varphi }_{q}}}} \right|,m=1,2,...,M
\end{equation}
Therefore, we obtain the suboptimal algorithm for the discrete model, i.e., solving the OP4. In fact, the algorithm for OP4 is the same as that for OP1, i.e., the Algorithm 3, except that right now, after the Algorithm 3 is convergent, we have to perform the projection operation via following (\ref{eq_dis}) to obtain the achievable discrete IRS reflecting coefficients.

\section{Simulation Analysis}\label{section_4}
In this section, the performance of the proposed algorithms are evaluated by numerical simulation. Considering two scenarios, they are, the strength of legitimate channel is superior or inferior to the eavesdropping channel. In the former case, the AP, legitimate user, eavesdropper and the IRS are located at (0, 0), (45, 0), (55, 0) and (50, 5) in meter (m) in a two-dimensional plane, respectively. And the latter exchanges the coordinates of the legitimate user with the eavesdropper. The other system parameters used in the simulations are following \cite{8641254} and \cite{7883924}, that is, we set the antenna number of all nodes as 4, namely, $N_T=N_R=N_E=4$. The noise power at both the legitimate receiver and the eavesdropper is set as $\sigma _R^2=\sigma _E^2=-40dBm$ and the maximum total transmitted power is set as $P_{\max}=30dBm$. Without loss of generality, all the channels are modeled as
%ÐŵÀÄ£ÐÍ
\begin{equation*}
\mathbf{H}=\sqrt{{\beta }/{\kappa +1}\;}\left( \sqrt{\kappa }{{\mathbf{H}}^{LoS}}+{{\mathbf{H}}^{NLoS}} \right),
\end{equation*}

\noindent where $\kappa$ is the Rician factor, while $\mathbf{H}^{LOS}$ and $\mathbf{H}^{NLOS}$ represent the deterministic line-of-sight (LoS) and Rayleigh fading/non-LoS (NLoS) components, respectively. $\beta$ represents the path loss, and is given by $\beta ={{\beta }_{0}}-10\alpha {{\log }_{10}}\left( {d}/{{{d}_{0}}}\; \right)$. Herein, $\beta _0$ denotes the path loss at the reference distance $d_0=1m$, $\alpha$ and $d$ represent the path loss exponent and the distance between the corresponding nodes, e.g., between AP and the IRS, or between AP and the legitimate user. Similarly to \cite{DBLP:journals/corr/abs-1907-12839}, we assume that the channels from IRS to legitimate user and the eavesdropper have LoS component and experience Rayleigh fading, simultaneously, however, the channels from AP to legitimate user, eavesdropper and IRS, only experience Rayleigh fading. Hence, the Rician factors are set as ${{\kappa}_{SR}}={{\kappa}_{SE}}=1$ and ${{\kappa }_{TR}}={{\kappa }_{TE}}={{\kappa}_{TS}}=0$. In addition, path loss exponents of all channels are set as ${{\alpha }_{TR}}={{\alpha }_{TE}}={{\alpha }_{TS}}={{\alpha}_{SR}}={{\alpha }_{SE}}=2$.

Furthermore, in order to better understand the positive effects of the IRS in improving the secure communication performance for the MIMOME system and the performance gain of the proposed algorithms, some benchmark schemes are introduced in the simulation based performance comparison and analysis, thus we have the following four algorithms, i.e., no-IRS, random-IRS, AO-based-IRS with $Q=2$ (or 8, 32) and the AO-based-IRS continues.

\textbf{\emph{no-IRS}:} That is, no IRS is used in the system and the secrecy rate is obtained by directly optimizing (9) under the conditions $\mathbf{G}_{TR}(\mathbf{\Theta}) = \mathbf{H}_{TR}$ and $\mathbf{G}_{TE}(\mathbf{\Theta}) = \mathbf{H}_{TE}$.

\textbf{\emph{random-IRS}:} That is, the reflecting coefficients of the IRS are randomly generated via following the rules that $|\theta_m|=1, m=1,..,M$ and $\theta_m$ follows an independent uniform distribution over $[0,2\pi)$.

\textbf{\emph{AO-based-IRS continues}:} It is our proposed alternative optimization based algorithm, i.e., the Algorithm 3.

\textbf{\emph{AO-based-IRS with $Q=2$ (or $8,$ $32$)}:} It is based on the Algorithm 3, except that right now, the reflecting coefficients can only take finitely discrete value and $Q$ determines the number of quantized reflection coefficient values of the IRS elements, i.e., $Q=2$, 8 or 32.

%·ÂÕæ½á¹ûͼ1disadvantaged
\begin{figure}[h]
    \centering
    \includegraphics[scale=0.8]{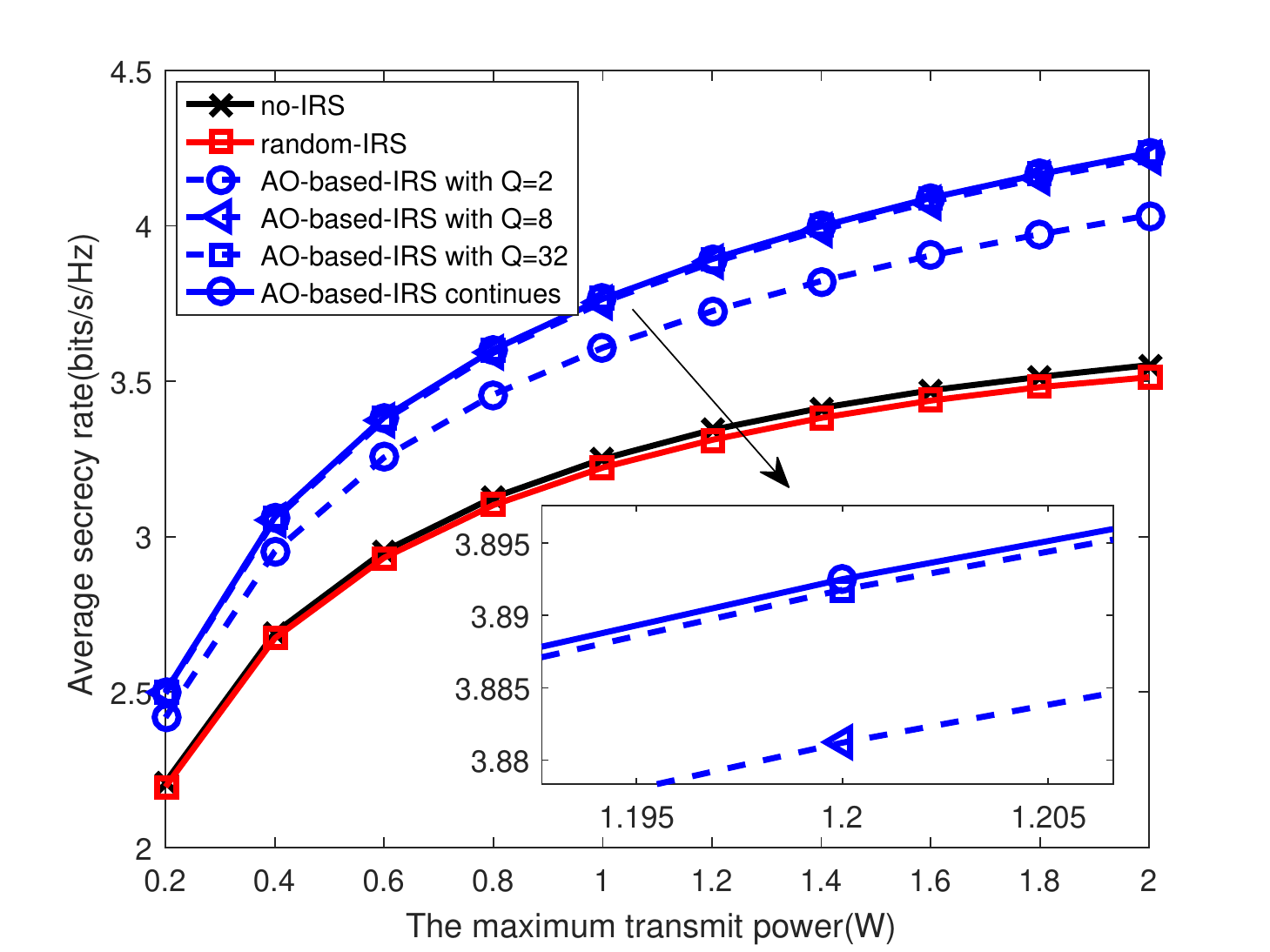}
    \caption{\quad Average secrecy rate VS $P_{max}$ (superior)}
    \label{Fig.2}
\end{figure}

At first, under the condition that the legitimate user channel (from AP to the legitimate user) is superior to the eavesdropper channel (from AP to the eavesdropper), the achievable secrecy rate of different algorithms are evaluated by varying the available transmission power constraint at the AP, i.e., the $P_{max}\in[0.2,2]W$ and $M=20$, and the result is shown in Fig. 2. One can note that, with the increase of the available transmission power at the AP, the achievable secrecy rates for all these algorithms are increased. This phenomenon is reasonable and it is consistent with the traditional MIMOME system \cite{DBLP:journals/tvt/CumananDSTL14, Steinwandt2014SECRECY, DBLP:journals/tsp/FakoorianS13a}. In addition, we note that among these algorithms, as expected, the algorithm `AO-based-IRS continues' obtains the best secrecy rate performance, then are those AO-based algorithm for IRS but with different quantization accuracy about the reflecting coefficients, i.e., `AO-based-IRS with $Q=2$', `AO-based-IRS with $Q=8$' and `AO-based-IRS with $Q=32$', and the algorithm `no-IRS' and `random-IRS' obtain the worst performance and are pretty close. Moreover, one can note that, on the one hand, with the increasing of the available transmission power at the AP, the performance gap between the algorithm `AO-based-IRS with $Q=x$' and the algorithm `AO-based-IRS continues' becomes larger, one the other hand, through increasing the quantization accuracy of the reflecting coefficients at the IRS, the performance gap between the algorithm `AO-based-IRS with $Q=x$' and the algorithm `AO-based-IRS continues' could be significantly reduced. It is indicated that taking $Q=8$ is sufficient for the system to obtain an acceptable secrecy rate with ignorable performance loss, i.e., less than $0.02bits/s/Hz$ even at $P_{max}=2W$, via comparing with the algorithm `AO-based-IRS continues'. Therefore, in the following, for the algorithm `AO-based-IRS with $Q=x$', we only consider the case $Q=8$.

\begin{figure}[h]
    \centering
    \includegraphics[scale=0.8]{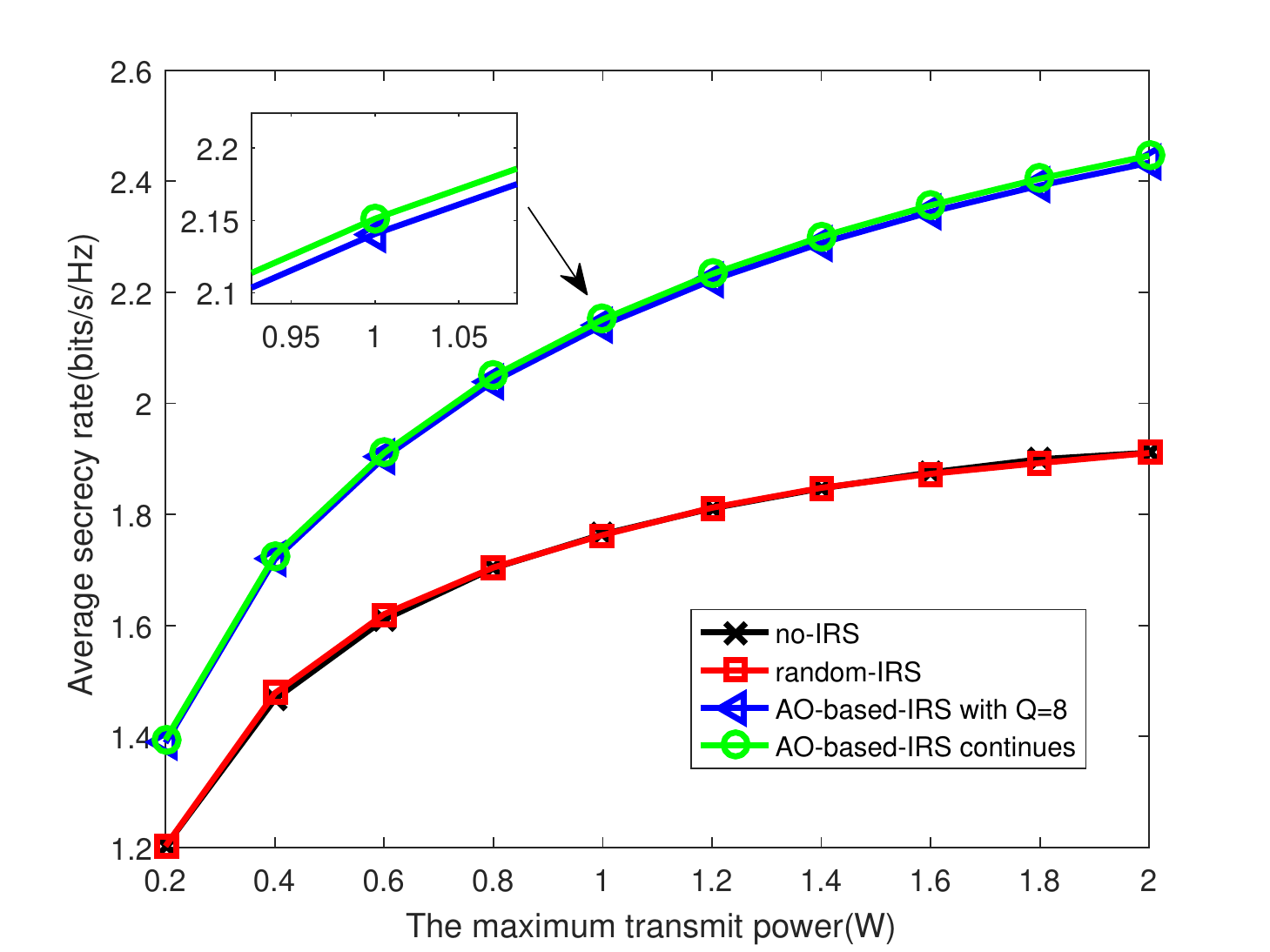}
    \caption{\quad Average secrecy rate VS $P_{max}$ (inferior)}
    \label{Fig.3}
\end{figure}

Then, with the same simulation parameters but under the condition that the legitimate channel is inferior to the eavesdropper channel, the achievable secrecy rate of different algorithms are evaluated again through changing the available transmission power constraint at the AP and the result is shown in Fig. 3. We can observe that, as expected, all the algorithms' obtained average secrecy rate is significantly reduced, however, the tendencies of the secrecy rate performance obtained by these algorithms are the same as that the legitimate channel is superior to the eavesdropper channel shown in Fig. 2, i.e., the algorithm `AO-based-IRS continues' obtains the largest secrecy rate, then it is the algorithm `AO-based-IRS with $Q=8$', and still the algorithm `no-IRS' and `random-IRS' have the similar performance with the least secrecy rate.

\begin{figure}[tbp]
    \centering
    \includegraphics[scale=0.8]{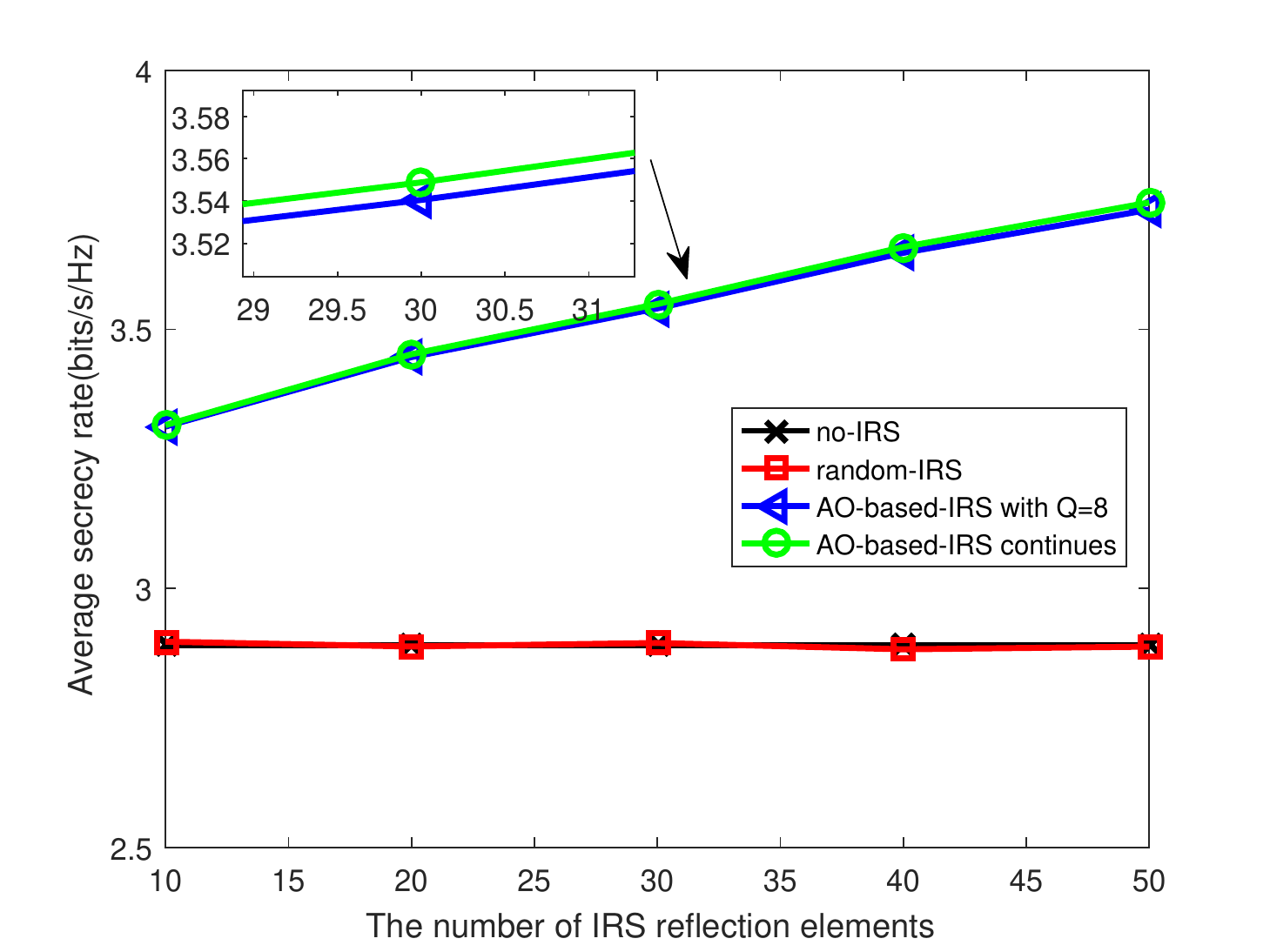}
    \caption{\quad Average secrecy rate VS number of IRS elements}
    \label{Fig.4}
\end{figure}

In Fig. 4, we further analyze how the system achievable secrecy rate performance is affected by the number of IRS elements in the system, i.e., from $M=10$ to $50$. Herein, still based on the condition that the legitimate channel is superior to the eavesdropper channel and the other simulation parameters are the same as that used in the Fig. 2. One can note that, for both the algorithm `AO-based-IRS with $Q=8$' and the algorithm `AO-based-IRS continues', their achievable secrecy rates are linear increment with the number of the IRS elements in the system. This increment comes from the factor more IRS elements in the system, more signal paths and energy could be reflected by the IRS to enhance the signal quality at the legitimate user but to reduce the signal quality at the eavesdropper. In addition, as expected, the performance of the algorithm `no-IRS' and `random-IRS' are not affected by the number of the IRS elements in the system.

\begin{figure}[h]
    \centering
    \includegraphics[scale=0.8]{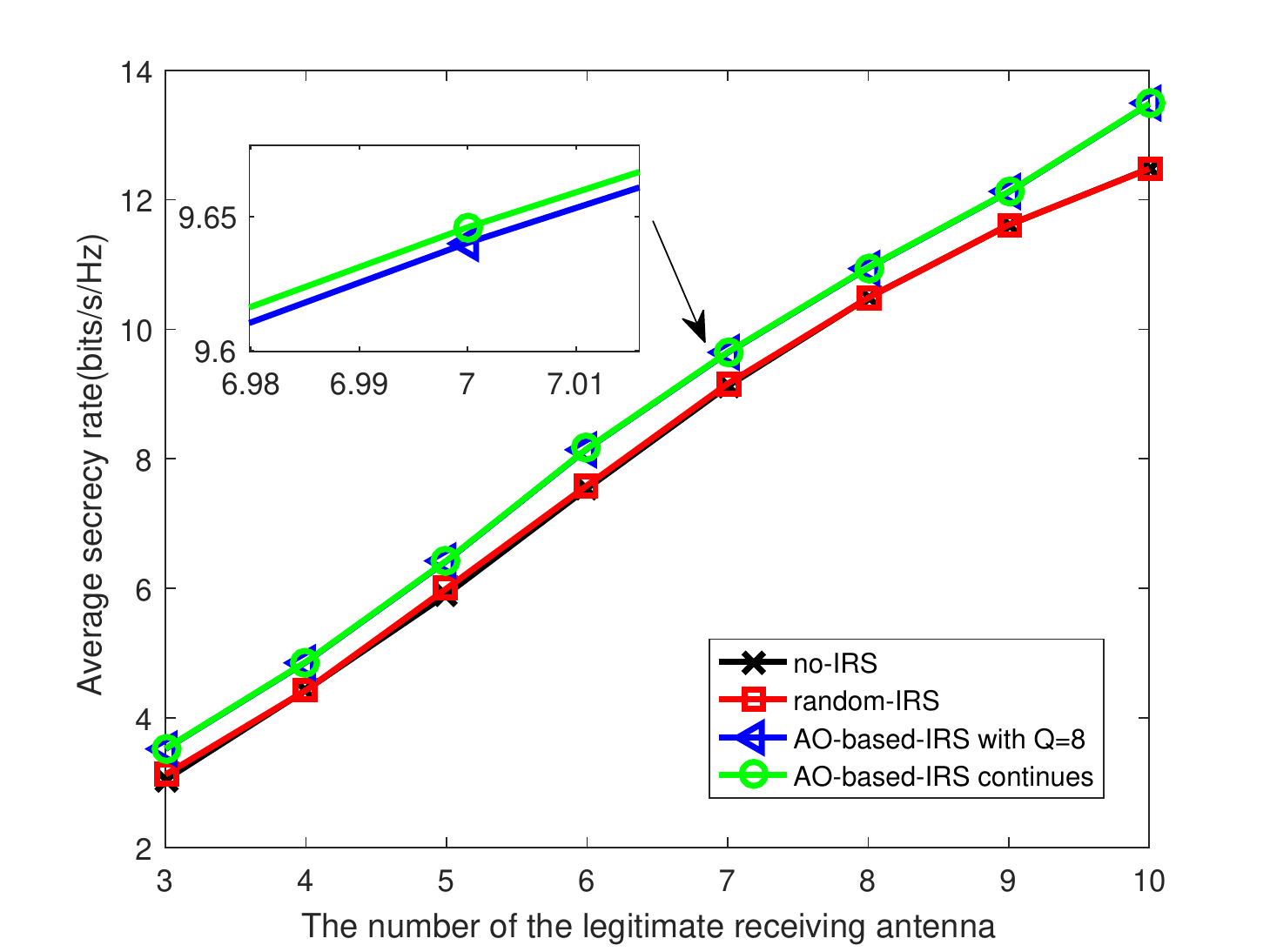}
    \caption{\quad Average secrecy rate VS number of legitimate receiving antennas}
    \label{Fig.5}
\end{figure}
Finally, in Fig. 5, the system achievable secrecy rate performance of the proposed algorithms are evaluated through changing the number of the legitimate receiving antennas, i.e., from $N_R=3$ to $10$, with given $N_T=10$ and $N_E=6$. As expected, with the increase of the number of the legitimate receiving antennas, the secrecy rate of all the algorithm are increased. This phenomenon comes from the following fact: the increase of the number of the legitimate receiving antennas can bring greater spatial diversity gain to the legitimate user, moreover, the rate of the legitimate user is increasing, as a result, the secrecy rate is increasing simultaneously. In addition, the performance of the algorithms `AO-based-IRS continues' and `AO-based-IRS with $Q=8$' is quite close, as well as the performance of the algorithms `no-IRS' and `random-IRS', which further verifies the above conclusion.

\section{Conclusion}
In this paper, the joint optimization of the transmit covariance matrix at the AP and the reflecting coefficients at the IRS to maximize the secrecy rate for the IRS-MIMOME system has been proposed and solved, with two different assumptions on the phase shifting capabilities at the IRS, i.e., the IRS has the continuous reflecting coefficients and the IRS has the discrete reflecting coefficients. For the former, due to the non-convexity of the formulated problem, an AO based algorithm has been proposed, i.e., for given the reflecting coefficients at the IRS, the SCA-based algorithm has been used to solve the transmit covariance matrix optimization, while given the transmit covariance matrix at the AP, alternative optimization has been used again in individually optimizing of each reflecting coefficient (i.e., $\theta_m$) at the IRS with the fixed of the other reflecting coefficients (i.e., $\theta_i,i=1,...,M,i\ne m$). For the individual reflecting coefficient optimization, the close-form or an interval of the optimal solution has been provided. Then, the overall algorithm has been extended to the discrete reflecting coefficient model at the IRS. Finally, from the numerical simulation evaluation, we have demonstrated that the proposed AO-based algorithm outperforms the other benchmark schemes. It has been also indicated that, for the IRS-MIMOME system, for practical IRS system with the discrete reflecting coefficient, taking $Q=8$ is sufficient for the system to experience less than $0.02bits/s/Hz$ performance loss even at $P_{max}= 2W$ and with more than $20$ IRS elements.

It is important to note that, in this paper, only one legitimate user and one eavesdropper are considered in the system, in the future, the more practical scenario with multiple legitimate users and multiple eavesdroppers would be concerned. In addition, in the system, the AP may use the AN to further improve the secrecy rate performance for the legitimate users. It is interesting and worth further studying the transmission strategies for these systems.

\appendices
\section{Proof of Proposition 3}\label{appendix_1}
%Here, we provide the proof of proposition 3.
Proof: Since $Tr\left(\mathbf{J}_R\right) \ne 0$ and $Tr\left(\mathbf{J}_E\right) = 0$, then the objective function in (\ref{eq22}) is transformed to (\ref{eq29}) and the problem is equivalent to maximize $\Re \left\{ {{\theta }_{m}}{{\lambda }_{E,m}} \right\}$ in (\ref{eq29}). In addition, $\Re \left\{ {{\theta }_{m}}{{\lambda }_{R,m}} \right\}\le \left| {{\theta }_{m}}{{\lambda }_{R,m}} \right|=\left| {{\theta }_{m}} \right|\left| {{\lambda }_{R,m}} \right|=\left| {{\lambda }_{R,m}} \right|$, and the inequality holds with equality if and only if ${{\hat{\theta }}_{m}}={{e}^{-j\arg \left( {{\lambda }_{R,m}} \right)}}$, thus we have the optimal solution conclusion for (\ref{eq22}). Furthermore, the corresponding optimal value can be obtained by substituting ${{\hat{\theta }}_{m}}={{e}^{-j\arg \left( {{\lambda }_{R,m}} \right)}}$ into (\ref{eq21}). That is, we have the proposition. $\hfill\blacksquare$

\section{Proof of Proposition 4}\label{appendix_2}
%The proof of proposition 4 is given here.
Proof: Since $Tr\left(\mathbf{J}_R\right)=0$ and $Tr\left(\mathbf{J}_E\right) \ne 0$, then the objective function in (\ref{eq22}) is transformed to (\ref{eq30}) and the problem is equivalent to minimize $\Re \left\{ {{\theta }_{m}}{{\lambda }_{E,m}} \right\}$ in (\ref{eq30}). Since $\Re \left\{ {{\theta }_{m}}{{\lambda }_{E,m}} \right\}\ge -\left| {{\theta }_{m}}{{\lambda }_{E,m}} \right|=-\left| {{\theta }_{m}} \right|\left| {{\lambda }_{E,m}} \right|=-\left| {{\lambda }_{E,m}} \right|$, where the inequality holds with equality if and only if ${{\hat{\theta }}_{m}}={{e}^{j\left( \pi -\arg \left( {{\lambda }_{E,m}} \right) \right)}}$, thus we have the optimal solution conclusion for (\ref{eq22}). Furthermore, the corresponding optimal value can be obtained by substituting ${{\hat{\theta }}_{m}}={{e}^{j\left( \pi -\arg \left( {{\lambda }_{E,m}} \right) \right)}}$ into (\ref{eq21}). Therefore, we have this proposition. $\hfill\blacksquare$

\section{Proof of Lemma 5}\label{appendix_3}
Proof: Herein, for the Lemma 5, we only present the proof of the conclusion (i) and the conclusion (ii) could be proved in the same manner thus it is omitted here for simplification. At first, we have
\begin{equation}
f(x)=\frac{a+b\cos x}{c+d\cos(x+\omega)}, x\in[0,2\pi)
\end{equation}

\noindent Since $a>b>0$, $c>d>0$ and $\omega\in[0,\pi)$, we know that $f_1(x) = a+b\cos x$ takes extremum at $x=0,\pi,2\pi$ and $f_2(x)=c+d\cos(x+\omega)$ takes extremum at $x=\pi-\omega, 2\pi-\omega$ over $x\in[0,2\pi)$. Following that, we can divided the definition domain of the function $f(x)$ into four regions, i.e., $D_1=[0,\pi-\omega]$, $D_2=(\pi-\omega,\pi)$, $D_3=[\pi,2\pi-\omega)$ and $D_4=[2\pi-\omega,2\pi)$. Then we prove that, for function $f(x),x\in[0,2\pi)$, it can take the maximization only at $x\in[0,\pi-\omega)$ if $\omega\in[0,\pi)$, that is, for $\forall x\in D_i,i=2,3,4$, $\exists{\hat{x}}\in D_1$ satisfies $f(\hat{x})>f(x)$. The proof is as follows.

\underline{Case $\forall x\in D_2$}: Let $\hat{x}_1 = \pi-\omega\in D_1$, then for $\omega\in[0,\pi)$ we always have $\cos x<\cos \hat{x}_1$ and $\cos(x+\omega)>\cos(\hat{x}_1+\omega)$, due to $a>b>0$ and $c>d>0$, then from the definition of $f(x)$ we know that $f(x)<f(\hat{x}_1)$.

\underline{Case $\forall x\in D_3 \bigcup D_4$}: Let $\hat{x}_2 = 2\pi-x \in [0,\pi]$, then for $\omega\in[0,\pi)$ we always have $\cos x=\cos \hat{x}_2$ and $\cos ( x+\omega )>\cos ( \hat{x}_2+\omega )$ due to $\cos x=\cos {\hat{x}_2}$ and $\sin x<\sin {\hat{x}_2}$; also since $a>b>0$ and $c>d>0$, then we have $f(x)<f(\hat{x}_2)$ from the definition of $f(x)$.

To sum up the case $\forall x\in D_2$ and case $\forall x\in D_3 \bigcup D_4$, we have the conclusion that the optimal solution $\hat{x} \in D_1$. Therefore, the conclusion (i) of the Lemma 5 is proved and in the same manner, we can also prove the conclusion (ii) of the Lemma 5. Finally, we have the Lemma 5. $\hfill\blacksquare$

\ifCLASSOPTIONcaptionsoff
  \newpage
\fi

\bibliographystyle{IEEEtran}
\bibliography{bibfile}

\end{document}